# Thermodynamics and kinetics of H adsorption and intercalation for graphene on *6H*-SiC(0001) from first-principles calculations

Yong Han*, James W. Evans, and Michael C. Tringides

Ames Laboratory, U.S. Department of Energy, Ames, IA 50011, United States
Department of Physics and Astronomy, Iowa State University, Ames, IA 50011, United States

Previous experimental observations for H intercalation under graphene on SiC surfaces motivate clarification of configuration stabilities and kinetic processes related to intercalation. From first-principles density-functional-theory (DFT) calculations, we analyze H adsorption and intercalation for graphene on a *6H*-SiC(0001) surface, where the system includes two single-atom-thick graphene layers: the top-layer graphene (TLG) and the underling buffer-layer graphene (BLG) above the terminal Si layer. Our chemical potential analysis shows that, in the low-H coverage regime (described by a single H atom within a sufficiently large supercell), intercalation into the gallery between TLG and BLG, or into the gallery underneath BLG, is more favorable thermodynamically than adsorption on top of TLG. However, intercalation into the gallery between TLG and BLG is most favorable. We obtain energy barriers of about 1.3 eV and 2.3 eV for a H atom diffusing on and under TLG, respectively. From an additional analysis of the energy landscape in the vicinity of a step on the TLG, we assess how readily one guest H atom on the TLG terrace can directly penetrate the TLG into the gallery between TLG and BLG versus crossing a TLG step to access the gallery. We also perform DFT calculations for higher H coverages revealing a shift in favorability to intercalation of H underneath BLG, as well as characterizing the variation with H coverage in interlayer spacings.

Email:
*y27h@ameslab.gov
ORCID iDs:
Yong Han https://orcid.org/0000-0001-5404-0911
James W. Evans https://orcid.org/0000-0002-5806-3720
Michael C. Tringides https://orcid.org/0000-0001-5003-7280

# I. INTRODUCTION

Systems involving graphene on a SiC surface are attractive given their potential for future applications to graphene-based electronics [1, 2]. Specifically, the system considered here includes two single-atom-thick graphene layers: the top-layer graphene (TLG) and the underling buffer-layer graphene (BLG) above the terminal Si layer of SiC. Intercalation by foreign guest atoms into few-layer van der Waals (vdW) materials (often called two-dimensional or 2D materials) including the graphene-SiC system has become a very active research area for developing next-generation energy-storage technologies and optoelectronic devices [3, 4, 5]. Since 2009, there have been many reports studying intercalation of guest atoms including H for the graphene-SiC system [6, 7, 8, 9, 10, 11, 12, 13, 14, 15, 16, 17, 18]. The initial motivation for H intercalation into the interface (or gallery) between BLG and terminal Si layer is to realize physical and electronic decoupling of the BLG from the SiC substrate, thereby obtaining quasi-freestanding graphene layer(s) [6, 7, 10, 15]. Studies of H intercalation with the graphene-SiC system have also led to interesting and important properties with potential applications, e.g., performance-enhanced graphene transistors [9] and quantum Hall effects [11] due to H intercalation, a possible route for engineering graphene-based devices [16], hydrogen storage [6, 7, 8], etc. In order to facilitate these promising applications, theoretical analysis of the thermodynamic stability of various configurations associated with H intercalation, and characterization of related kinetic processes, are important. However, such theoretical analysis in literature is lacking.

In a broader context, there is much current interest in intercalation of metals (in addition to H), not just underneath graphene on SiC, but also underneath graphene on metal substrates [19], and underneath the graphene surface layer of graphite [20]. With regard to the latter, in recent years, Thiel Research Group with collaborators at Iowa State University have performed experimental and theoretical studies of the intercalation of multiple metals (e.g., Dy, Cu, Ru, Fe, Ag, Au, Pt, etc.) into the near-surface region of graphite [20, 21, 22, 23, 24, 25, 26, 27, 28, 29, 30]. They demonstrated that it is difficult or impossible to directly penetrate the perfect TLG of graphite into the galleries beneath the top graphene layer for temperatures below about 1500 K by deposition, even if followed by annealing in an ultrahigh vacuum environment. This feature generally implies a large energy barrier for penetration of the guest atom. Thus, to effectively intercalate metals into graphite, an ion-bombarded graphite surface was always used, so that the guest metal atoms (e.g., Dy, Cu, Ru, Fe, and Pt) could intercalate through portal defects created by ion-bombardment [20, 21, 22, 23, 24, 25, 26, 27, 28, 29, 30].

However, for epitaxial graphene on a SiC substrate, previous experiments show that many types of guest elements including H can intercalate without a pre-prepared ion-bombarded top surface [31]. There is still a lack of a comprehensive analysis on how guest atoms are inserted under the top graphene layer of the graphene-SiC system. Therefore, analysis of the relevant kinetic process for intercalation of a given guest atom in this system is needed.

Sufficiently extensive first-principles density functional theory (DFT) calculations analyzing the thermodynamic stabilities and geometries of various adsorbed and intercalated guest atom configurations, as well as kinetic barriers for intercalation in the graphene-SiC system, have the potential to elucidate intercalation pathways. A recent such analysis for the rare-earth metal Dy demonstrated that direct penetration of (perfect) TLG is difficult, while more facile intercalation can occur by crossing specific TLG steps [32]. In the present work, we perform DFT calculations for H as the guest atom intercalating under graphene on a 6$H$-SiC(0001) substrate. By calculating the chemical potentials of H at different positions, we determine the thermodynamic favorability of various adsorbed versus intercalated configurations. To assess relevant kinetics, we determine the binding energy landscape, and thus diffusion barriers, for the transition of a single H atom from top terrace to different types of TLG steps and then to the gallery underneath the TLG.

This paper is organized as follows. Sec. II briefly describes the DFT method used in this work, and the model used for a clean graphene-SiC system without guest H atoms. In Sec. III and Sec. IV, we present our DFT results in the low H-coverage regime (described by a single H atom within a sufficiently large supercell) for thermodynamics as well as diffusion and direct intercalation barriers for the graphene-SiC system without and with TLG steps, respectively. In Sec. V, we discuss the binding energy landscapes and intercalation kinetics of a single H atom via TLG steps. Further discussion is provided in Sec. VI which also draws upon additional DFT results in the Appendix for higher H coverages, providing information on the coverage dependence of the chemical potentials of H and thus on intercalation preferences. In Sec. VII, we provide a summary. Additional information is provided in Supplementary Material [33].

**II. DFT METHODOLOGY AND MODELS FOR GRAPHENE-SiC SYSTEM**

Application of the DFT method to analyze the graphene-SiC system, as implemented here, has been described previously [32]. Briefly, we use the VASP code [34] with the projector-augmented-wave pseudopotentials [35] and the optB88-vdW functional [36]. The method has already been proven very successful when applied for various vdW materials with metals [22, 23, 37, 25, 24, 26, 27, 38, 28, 29, 20]. Benchmark analyses of bulk properties of 6$H$-SiC, graphene, and graphite from this method closely match experimental data [37, 25, 32]. Also [39], using this DFT method for a $H_2$ molecule in gas phase, we obtained a reliable bond length of 0.7442 Å which in good agreement with the experimental value of 0.74130 Å [40], and a bond strength of 4.989 eV which is reasonably consistent with the experimental value of 4.47718 eV [41].

As described in Sec. I, our system involves a guest H atom plus TLG on BLG supported by a Si-terminated 6$H$-SiC(0001) substrate. According to previous experimental observations, graphene on the SiC substrate can display a $(6\sqrt{3} \times 6\sqrt{3})R30°$ superlattice ordering [42, 43] which evolves at lower temperatures from superstructures including the $(\sqrt{3} \times \sqrt{3})R30°$ superlattice ordering [42]. Correspondingly, there are two models can be used to construct the unit cells for DFT calculations [44, 45, 15, 46, 47, 48, 32, 39]. In Model 1, $13a_C^* \times 13a_C^*$ matches $6\sqrt{3}a_{SiC} \times 6\sqrt{3}a_{SiC}$, and in Model 2, $2a_C^* \times 2a_C^*$ matches $\sqrt{3}a_{SiC} \times \sqrt{3}a_{SiC}$, where $a_{SiC}$ is the lateral lattice constants of SiC(0001) and $a_C^*$ is the strained overlayer graphene lattice constant. If we take our DFT values $a_{SiC} = 3.09545$ Å [32] and the unstrained graphene lattice constant $a_C = 2.464$ Å [37], there is a tiny laterally tensile strain $(a_C^*/a_C - 1) \times 100\% \approx 0.4\%$ with $a_C^* = 6\sqrt{3}a_{SiC}/13 \approx 2.475$ Å for Model 1 and a relatively larger lateral tensile strain $(a_C^*/a_C - 1) \times 100\% \approx 8.8\%$ with $a_C^* = \sqrt{3}a_{SiC}/2 \approx 2.681$ Å for Model 2. For each DFT calculation below, we use a rhombohedral supercell with the size of $2ma_C^* \times 2na_C^*$ matching $m\sqrt{3}a_{SiC} \times n\sqrt{3}a_{SiC}$, or a rectangular supercell with the size of $2ma_C^* \times n\sqrt{3}a_C^*$ matching $m\sqrt{3}a_{SiC} \times 1.5na_{SiC}$ (or $n\sqrt{3}a_C^* \times 2ma_C^*$ matching $1.5na_{SiC} \times m\sqrt{3}a_{SiC}$), where $m$ and $n$ are positive integers.

In this work, we do not use Model 1 because of its large computational cost [32]. Instead, we use Model 2 for all DFT calculations, noting that it has already been widely utilized in literature [44, 45, 15, 46, 47]. Strain effects are not expected to affect the main conclusions [32, 39]. To obtain the energy barriers for various diffusion processes involving a H atom, we use the climbing image nudged elastic band (CINEB) method [49] to calculate the minimum energy paths (MEPs). For all structure optimizations in this work, the convergence criterion of total energy is that the force exerted on each relaxed atom is less than 0.01 eV/Å. In addition, the nuclear quantum effects (including zero-point motion, discrete vibrational levels, and tunneling) for lightweight H are not considered in the calculations of this work because these effects are expected to be insignificant at an experimentally typical intercalation temperature (far above 100 K) [39].

## III. ADSORPTION, INTERCALATION, AND DIFFUSION OF A SINGLE H ATOM WITHOUT TLG STEPS

### A. Chemical Potentials

Characterizing the thermodynamic preference of a guest atom for intercalation into versus adsorption on the surface of a layered material [19, 20], as well as related kinetic processes [21, 30], is based on analysis of the chemical potential, $\mu$, for a single guest H atom within a sufficiently large supercell, i.e., in the low-H coverage regime. For a system with a single H atom in the supercell, one has

$$\mu = E_{tot} - E_{cln} - E_H, \tag{1}$$

where $E_{tot}$ is the total energy of the H structure plus graphene and SiC, $E_{cln}$ is the energy of the fully relaxed clean graphene-SiC system (without guest H) with or without steps, and $E_H$ is the energy of one isolated H atom in vacuum. In cases where we consider a system with a TLG step saturated with H atoms plus one additional guest H atom, it is more instructive to consider the binding energy $E_{bind}$ of that guest atom to the rest of the system. $E_{bind}$ corresponds to $\mu$ for a system with a single guest H atom. See Sec. S1 for further discussion [33]. Prediction of basic geometric structural parameters after the intercalation into the layered material can be also instructive, e.g., interlayer spacings and surface corrugations, which can be measured in experiments [13, 15].

Previously, we used Model 2 to construct a rhombohedral supercell with $m = n = 2$ for the graphene-SiC system with a SiC substrate of 6 single-atom-thick layers [32], as shown in Figs. 1(a) and 1(b). After full relaxation, the interlayer spacings from our optB88-vdW calculations are almost the same as those from previous PBE-TS calculations [15], but the corrugations are significantly smaller than the PBE-TS calculations, as listed in Table S1 [33]. The interlayer spacings and the corrugations are defined in the caption of Fig. 1. Strictly, it is not possible to analyze an isolated atom with periodic boundary conditions where the supercell is finite. In this work, we choose the supercell in Fig. 1(a), as indicated by the green frame. The $k$ mesh is taken to be $6 \times 6 \times 1$. In this section, only one H atom is included within the supercell, equivalent to a coverage of 1/16 ML, where 1 ML has the definition that any $a_C^* \times a_C^*$ unit cell for graphene is occupied by one H atom to form a H monatomic layer (ML) [32]. We expect that the coverage of 1/16 ML is sufficiently low to provide a good approximation to the description of a single isolated H atom.

To search for the energy minima of the H atom on TLG or in a gallery, we first relax $N = 9$ configurations with the initial positions of the H atom at the 9 green crosses in Fig. 1(a). By comparing total energies and examining geometries after relaxation, we can find the fully relaxed configurations with energy minima. In principle, to obtain all possible local minima, a larger $N$ is required. However, to find the global minimum on TLG or in a gallery, the calculations from the above $N = 9$ is sufficient. Also note that the global minimum can be verified from the selected CINEB calculations, as analyzed below.

In Fig. 1(c), we plot the chemical potentials $\mu$ of the 27 fully relaxed structures for a guest H atom adsorbed on TLG (configuration a1 to a9), intercalated into BT gallery (b1 to b9), and intercalated into BS gallery (c1 to c9). See Table S2 for the corresponding values [33]. The initial positions of the H atom for these configurations are indicated in Fig. 1(a) by 9 green crosses (labelled as $i = 1, 2, \ldots$ and 9) in a $\sqrt{3}a_{\text{SiC}} \times \sqrt{3}a_{\text{SiC}}$ unit cell and in Fig. 1(b) for a$i$, b$i$, and c$i$, correspondingly from top to BT to BS gallery, respectively.

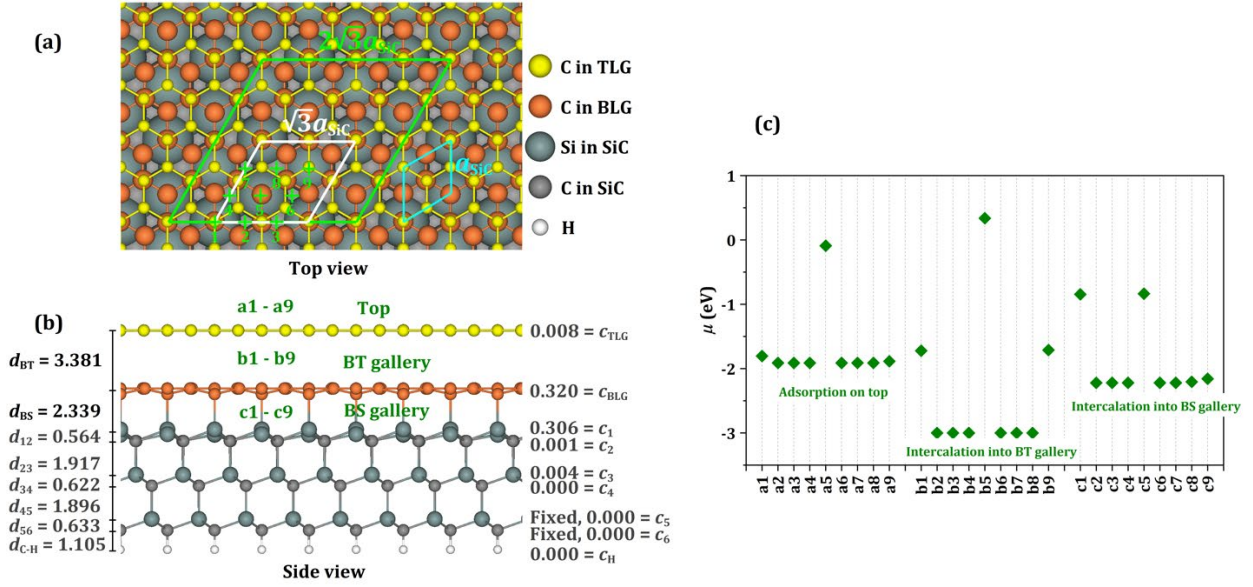

**FIG. 1.** (a) Top view and (b) side view of fully relaxed graphene layers (TLG and BLG) on a Si-terminated 6H-SiC(0001) substrate from our DFT calculation. The SiC substrate has a thickness of 6 single-atom-thick layers with a C-Si-C-Si-C-Si stacking sequence from bottom to top. The bottommost C and Si single-atom-thick layers are fixed during relaxation, and the dangling bonds of C atoms are passivated by pseudo H atoms. The green frame in top view indicates a supercell ($4a_C^* \times 4a_C^*$ matching $2\sqrt{3}a_{SiC} \times 2\sqrt{3}a_{SiC}$) used to the DFT calculation. The white and cyan frames in top view indicate a rotation of 30° from $a_{SiC} \times a_{SiC}$ unit cell to $\sqrt{3}a_{SiC} \times \sqrt{3}a_{SiC}$ unit cell. The BT gallery is between BLG and TLG, while the BS gallery is between BLG and terminal Si layer. The left of side view shows the interlayer spacings ($d_{BT}$, $d_{BS}$, $d_{12}$, $d_{23}$, $d_{34}$, $d_{45}$, $d_{56}$, and $d_{C-H}$), each of which is defined as the difference of average heights of atoms between two corresponding single-atom-thick layers. The right of side view shows the corrugations ($c_{TL}$, $c_{BL}$, $c_1$, $c_2$, $c_3$, $c_4$, $c_5$, $c_6$, and $c_H$), each of which is defined as the height difference between the highest atom and the lowest atom within the corresponding single-atom-thick layer. All interlayer spacings and the corrugations are in units of Å. (c) The chemical potentials ($\mu$) of fully relaxed structures for a guest H atom adsorbed on TLG (configuration a1 to a9), intercalated into BT gallery (b1 to b9), and intercalated into BS gallery (c1 to c9). The initial positions of the H atom for these configurations are indicated in the top view (a) by 9 green crosses in a $\sqrt{3}a_{SiC} \times \sqrt{3}a_{SiC}$ unit cell. Initially, the position of the guest H atom has a shift from a1 to b1 or c1, from a2 to b2 or c2, …, and from a9 to b9 or c9 (i.e., correspondingly from top to BT or BS gallery), as indicated in (b).

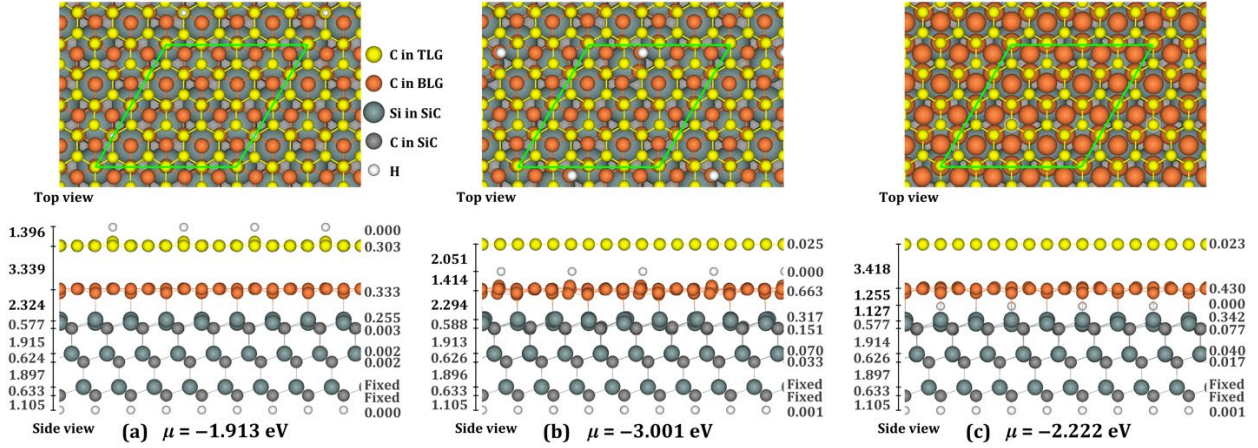

**FIG. 2.** Fully relaxed structure with the lowest $\mu$ value for a guest H atom (a) on TLG, (b) in BT gallery, and (c) in BS gallery from our DFT calculations, which correspond to a4, b3, and c6 in Fig. 1, respectively. The green frame in top view indicates a $2\sqrt{3}a_{SiC} \times 2\sqrt{3}a_{SiC}$ supercell used to the DFT calculations. The left of side view shows the interlayer spacings (in Å) and the right shows the corrugations (in Å) for each single-atom-thick layer (for definitions of interlayer spacings and corrugations, see the caption of Fig. 1).

In Fig. 2, we show three configurations a4, b3, and c6, each of which corresponds to the lowest of the nine $\mu$ values for the H atom on TLG, in BT gallery, or in BS gallery, with the chemical potentials $\mu = -1.913$ eV, $-3.001$ eV, and $-2.222$ eV, respectively. Also note that six configurations a$i$, b$i$, or c$i$, $i = 2, 3, 4, 6, 7$, and 8 have almost the same lowest energies [Fig. 1(c)]. These $\mu$ values show that intercalation into BT gallery is most favorable, while the adsorption on TLG is most unfavorable, i.e., the order for the thermodynamic favorability of the H atom is BT > BS > top corresponding to the chemical potentials from low to high. Here, it is interesting to mention that, from our previous DFT calculations for freestanding bilayer graphene, intercalation of a H atom into the gallery is less favorable than adsorption on top [29, 39]. This indicates that the SiC substrate can significantly change the preferred position of the H atom on top or in the gallery.

The interlayer spacings and corrugations of all layers (defined in the caption of Fig. 1) are provided for the configurations in Fig. 2 and also listed in Table S1 for $d_{BT}$ and $d_{BS}$. Comparing the interlayer spacing for any BT or BS gallery *with* and *without* an intercalated H atom reveals only a tiny change less than 2.5%. For the corrugations, there are variations for different layers. Overall, the C or Si layers closer to the guest H layer have relatively larger corrugations from 0.25 to 0.70 Å, as indicated in the side views in Fig. 2.

**B. Diffusion and Direct Penetration Barriers**

To obtain energy barriers for a single H atom diffusing on TLG and in the BT gallery, as well as to obtain the barrier for penetrating TLG into the BT gallery, we determine the corresponding MEPs. For the diffusion on TLG, two CINEB endpoints e1 and e2 are chosen, corresponding to two adjacent configurations a3 and a4 in Fig. 1, respectively. The MEP and trajectory from our DFT calculations is plotted in Fig. 3(a), and then the corresponding diffusion barrier $E_d^{top} = 1.31$ eV is obtained, as indicated by a vertical red arrow. For diffusion in the BT gallery, we find an MEP and the trajectory between e1 and e2 (corresponding to b4 and b3 in Fig. 1, respectively) with a diffusion barrier of 3.05 eV, as indicated by a vertical red arrow in Fig. 3(b). This barrier is significantly larger than 2.29 eV, which was previously obtained from another MEP in Fig. 3(c) [39]. The two endpoints e1 and e2 in Fig. 3(c) correspond to m2 and m3 in Fig. 3 of Ref. [39], where m2 (or e2) corresponds to b6 in Fig. 1, while m3 corresponds to the H atom at a site below and close to a C atom of TLG, as illustrated by the trajectory. Thus, we take the diffusion barrier in BT gallery to be $E_d^{BT} = 2.29$ eV, i.e., the more favorable diffusion path for the H atom in BT gallery is the MEP in Fig. 3(c) but not that in Fig. 3(b). For the penetration from on top of TLG into the BT gallery, we previously obtained a global energy barrier is $E_{dp} \approx 1.8$ eV. DFT calculations for the penetration process are very demanding, and the details as well as discussion have been separately published [39]. One caveat, as discussed further below, is that this latter result is impacted by the significant tensile strain of about 8.8% in Model 2, and that other models with lower strain plausibly give higher values for $E_{dp}$.

For the diffusion of a metal atom on top of or in the gallery of a layered vdW material, the diffusion barriers are generally significantly less than 1 eV, e.g., about 0.01 to 0.02 eV for a Cu-graphite system [25], about 0.20 to 0.30 eV for a Cu-MoS$_2$ system [38], and about 0.4 to 0.6 eV for a Dy-graphene-SiC system [32]. In contrast, for a H atom, the bonding with an adjacent C atom is a covalent-like short-range interaction [29]. Thus, the diffusion of the H atom from one site near a C atom to a site near an adjacent C atom is equivalent to breaking the strong covalent-like bond with the first C atom and then recombining with the second C atom. This results in the significantly larger diffusion barriers $E_d^{top} = 1.31$ eV on TLG and $E_d^{BT} = 2.29$ eV in BT gallery for a H atom than for a metal atom, as listed above. The barrier $E_d^{BT} = 2.29$ eV for the H atom in the BT gallery is even larger than the corresponding penetration barrier $E_{dp} \approx 1.8$ eV, because the C-H bond is not broken for the penetration during which the H atom moves only around a TLG C atom [39].

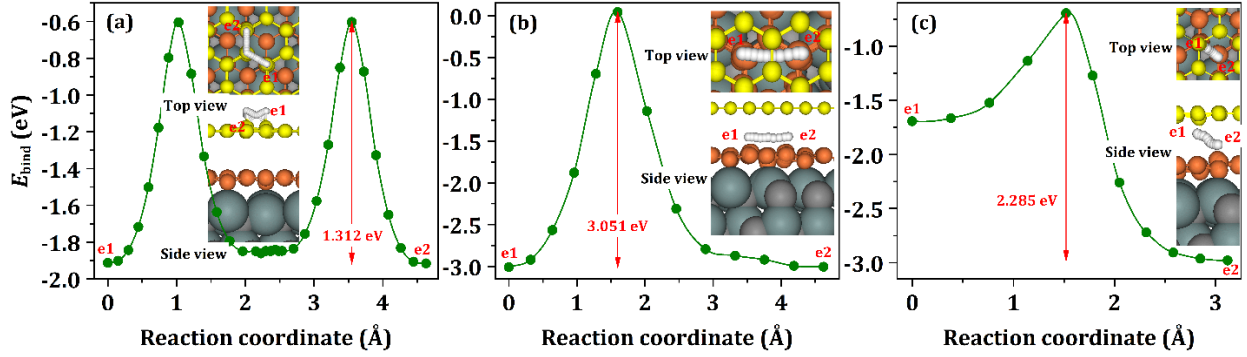

**FIG. 3.** MEPs between two endpoints e1 and e2 for a single H atom diffusing (a) on TLG with e1 and e2 corresponding to a3 and a4 in Fig. 1, (b) in BT gallery with e1 and e2 corresponding to b4 and b3 in Fig. 1, and (c) in BT gallery with e1 and e2 corresponding to m2 and m3 in Fig. 3 of Ref. [39] (and e2 also corresponding to b6 in Fig. 1) from our DFT calculations. The insets are top and side views of trajectories from one endpoint to another one. Green dots are the CINEB images, and the curves are from nonlinear interpolations. The diffusion barriers are indicated by the vertical red arrows.

## IV. ADSORPTION AND INTERCALATION OF A SINGLE H ATOM AT TLG STEPS

Motivated by previous experiments, in which domain boundaries between intrinsic stacking domains in TLG are observed [18, 50, 51, 52], we also consider whether a TLG step can act as a significant pathway or portal for H intercalation. Such analysis is analogous to that in our previous work for the Cu-graphite system [25] and the Dy-graphene-SiC system [32]. For our DFT calculations in this section, the system includes one guest H atom plus the graphene-SiC system with stepped TLG (i.e., with incomplete TLG strips or ribbons in the supercell, and exposed BLG strips separated from the TLG ribbons by linear steps). Our calculations will involve three typical step edge structures for the TLG: zigzag, armchair, and zz57 (i.e., a reconstruction of zigzag step edge by a transition from a $C_6$-ring sequence to a $C_5$-ring-plus-$C_7$-ring sequence). The supercells and corresponding $k$ meshes in the calculations below are the same as those chosen in previous work [32].

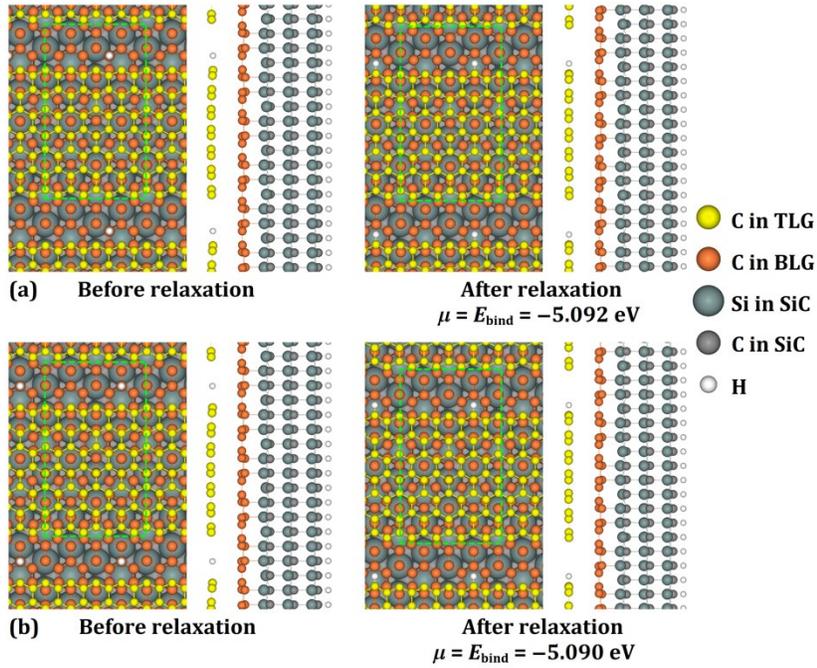

**FIG. 4.** Top and side views of structures for zz-1 steps with one H atom in the $2\sqrt{3}a_{SiC} \times 6a_{SiC}$ supercell (green dashed frame) used in our DFT calculations. The zz-1 step is initially obtained by removing one zigzag row and then relaxing. Initial (before relaxation) and final fully relaxed (after relaxation) configurations are indicated. Note that the initial positions of the H atom in (a) and (b) are different. During energy minimization, except all relaxed atoms, the bottommost C and Si single-atom-thick layers are fixed. Within the supercell, we also fix the lateral coordinates of four C atoms of the step edge without any H atom. The chemical potentials ($\mu$) and binding energies ($E_{bind}$) of configurations after relaxation are indicated.

## A. TLG zigzag steps

For the structures of the TLG zigzag steps, we consider three configurations. The first configuration is zz-1, for which one zigzag C row is removed so that a narrow strip of exposed BLG is created with two zigzag steps within a supercell, as shown in Fig. 4. The second configuration is zz-3, for which three zigzag rows are removed so that a wider strip of exposed BLG is created with two zigzag steps within a supercell, as shown in Fig. 5. The strip of exposed BLG for zz-3 is sufficiently wide so that the interactions between two steps can be neglected and then one of the steps can be approximated as an isolated zigzag step. Thus, zz-1 and zz-3 correspond to two limits: two steps separated by a narrowest strip of exposed BLG and one step approximately isolated, respectively.

For the zz-1 calculations, we initially position the H atom at two differently selected sites near the zz-1 step edge in Figs. 4(a) and 4(b), and then obtain two fully relaxed configurations. Because these two fully relaxed configurations are almost identical in geometry and their chemical potentials $\mu$ or binding energies $E_{bind}$ have a tiny difference of $\Delta\mu = \Delta E_{bind} = -5.090 - (-5.092) = 0.002$ eV, we view them as the same configuration. See Sec. S1 [33] for the definition of $E_{bind}$ as well as the relationship between $\mu$ and $E_{bind}$.

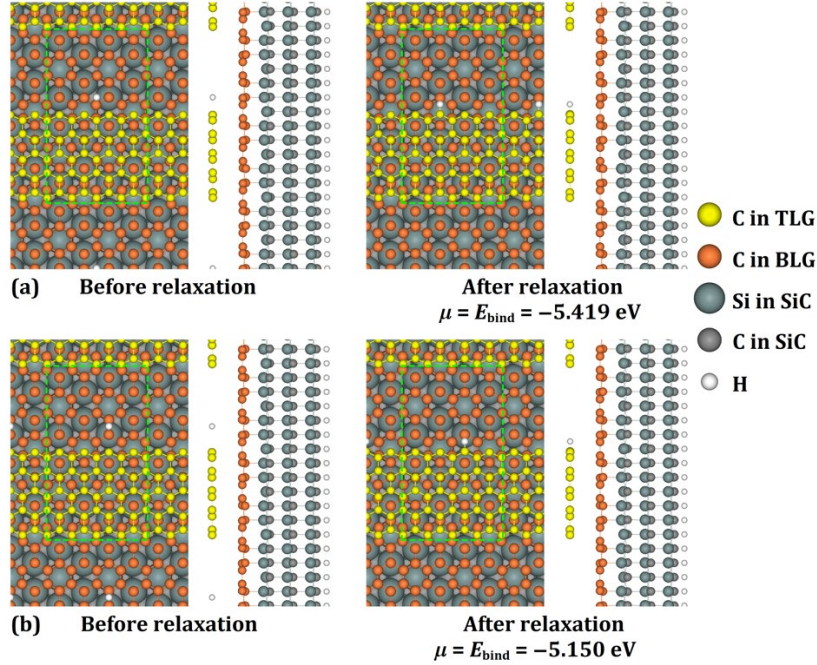

**FIG. 5.** Top and side views of structures for zz-3 steps with one H atom in the $2\sqrt{3}a_{SiC} \times 6a_{SiC}$ supercell (green dashed frame) used in our DFT calculations. The zz-3 step is initially obtained by removing three zigzag rows and then relaxing. Note that the initial positions of the H atom in (a) and (b) are different. Other details are similar to the caption in Fig. 4.

For the zz-3 calculations, we initially position the H atom at two differently selected sites near the zz-3 step edge in Figs. 5(a) and 5(b), and then obtain two fully relaxed configurations, which have different geometries and have a chemical-potential or binding-energy difference of $\Delta\mu = \Delta E_{bind} = -5.150 - (-5.419) = 0.269$ eV. Thus, we view the fully relaxed configuration in Fig. 5(a) as the global energy minimum for the zz-3 step. Also note that, after full relaxation, the H atoms in Figs. 5(a) and 5(b) have similar positions relative to graphene, but the positions are different relative to the Si termination, i.e., the position of the H atom relative to the terminal Si layer has a shift from Fig. 5(a) to Fig. 5(b). This indicates a substrate effect resulting in the above chemical potential difference of 0.269 eV.

## B. TLG armchair steps

For the TLG armchair step structures, we consider only two configurations. The first configuration is ac-1, for which one armchair row is removed so that a narrow strip of exposed BLG is created with two armchair steps within the supercell, as shown in Fig. 6. The second configuration is ac-3, for which three armchair rows are removed so that a wider strip of exposed BLG is created with two armchair steps, as shown in Fig. 7. Again, the strip of exposed BLG for ac-3 is sufficiently wide so that the interactions between two steps can be neglected and then one of the steps can be approximated as an isolated armchair step.

For the ac-1 calculations, we initially select two different sites of the H atom near the ac-1 step edge in Figs. 8(a) and 8(b), and then obtain two fully relaxed configurations. From the symmetry of the system with periodic boundary conditions, these two fully relaxed configurations are nearly identical. Their chemical potentials $\mu$ or binding energies $E_{bind}$ have a small difference of $\Delta \mu = \Delta E_{bind} = -3.666 - (-3.672) = 0.006$ eV, likely because we fix the lateral coordinates of four C atoms of the step edge initially without any H atom. Neglecting this small energy difference, we can view any one of them as the global energy minimum for the ac-1 step.

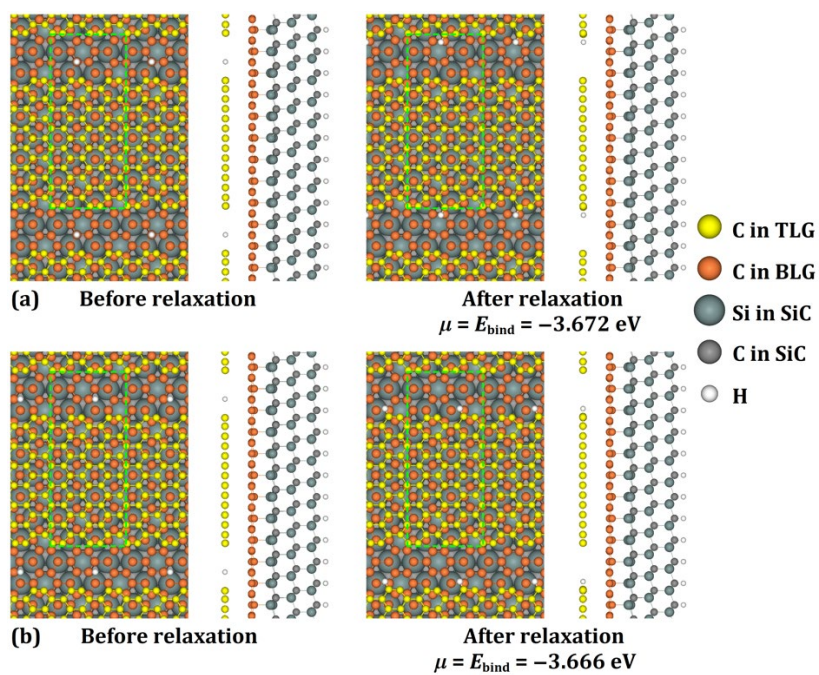

**FIG. 6.** Top and side views of structures for ac-1 steps with one H atom in the $3a_{SiC} \times 4\sqrt{3}a_{SiC}$ supercell (green dashed frame) used in our DFT calculations. The ac-1 step is initially obtained by removing one armchair row and then relaxing. Note that the initial positions of the H atom in (a) and (b) are different. Other details are similar to the caption in Fig. 4.

For the ac-3 calculations, we also initially select two different sites of the H atom near the ac-3 step edge in Fig. 7(a) and 7(b), and then obtain two fully relaxed configurations which are very different in geometry. The configuration in Fig. 7(b) has a much lower chemical potential or binding energy $\mu = E_{\text{bind}} = -3.673$ eV than $\mu = E_{\text{bind}} = -0.116$ eV for that in Fig. 7(a), and therefore is viewed as the global energy minimum for the ac-3 step. This result indicates that the H atom with a closer C atom at the edge has a much stronger bond strength than with farther C atoms by comparing the position of the H atom in Fig. 7(b) with Fig. 7(a) after full relaxation. This, again, reflects the strong covenant-like short-range C-H interaction.

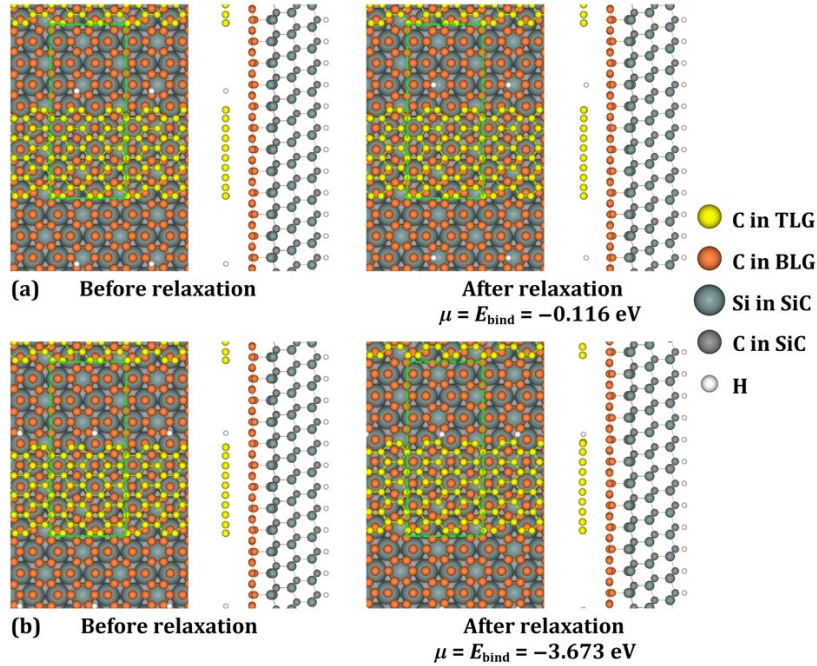

**FIG. 7.** Top and side views of structures for ac-3 steps with one H atom in the $3a_{\text{SiC}} \times 4\sqrt{3}a_{\text{SiC}}$ supercell (green dashed frame) used in our DFT calculations. The ac-3 step is initially obtained by removing three armchair rows and then relaxing. Note that the initial positions of the H atom in (a) and (b) are different. Other details are similar to the caption in Fig. 4.

### C. TLG zz57 steps

For the zz57 step calculations, we only consider the zz57-3 step which is obtained by modifying a zz-3 step in Fig. 5. We initially select three different sites of the H atom near the zz57-3 step edge in Figs. 8(a), 8(b), and 8(c), and then obtain three fully relaxed configurations with $\mu = E_{\text{bind}} = -0.171$ eV, $-3.068$ eV, and $-2.906$ eV, respectively. For Fig. 8(c), the initial position of the H atom is chosen to be beneath the TLG ribbon, but that for Fig. 8(a) or 8(b) is not. The fully relaxed configuration in Fig. 8(b) has the lowest energy and therefore is viewed as the global energy minimum for the zz57-3 step.

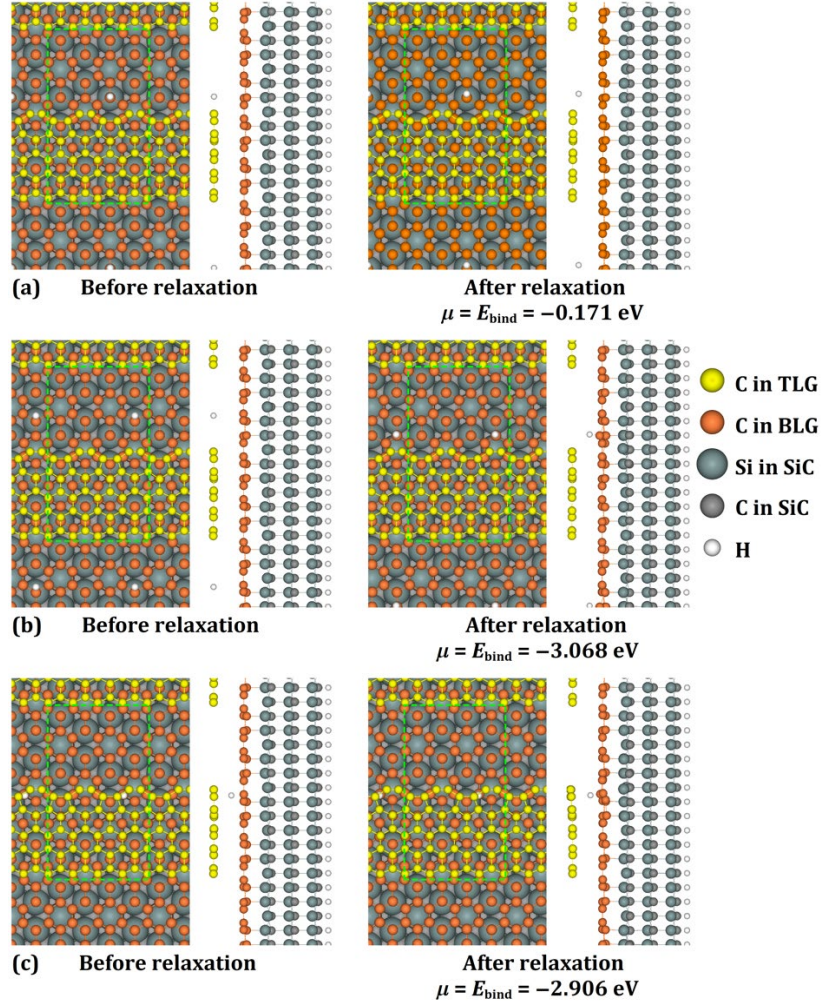

**FIG. 8.** Top and side views of structures for one H atom at zz57-3 step edges in the $2\sqrt{3}a_{SiC} \times 6a_{SiC}$ supercell (green dashed frame) used in our DFT calculations. The zz57-3 step is initially constructed by alternatively connecting C$_5$ and C$_7$ rings [25] from modifying a zz-3 step and relaxing. Note that the initial positions of the H atom in (a), (b), and (c) are different. Other details are similar to the caption in Fig. 4.

### D. H-saturated TLG zigzag steps

All of the above cases consider an isolated H atom at clean steps. However, plausibly, steps are quickly saturated with H after deposition commences. This can dramatically change the energetics of an additional H atom in the vicinity of the step, and potentially create more viable pathways for intercalation. We consider one such case denoted by zz-3-c, for which a zz-3 step edge is pre-decorated by a H chain, as shown in Fig. 6.

For the zz-3-c calculations, we initially position the H atom at two different sites near the H chain in Figs. 6(a) and 6(b). After full relaxations, we obtain configurations with $E_{bind} = -0.147$ eV and $-3.041$ eV, respectively. Thus, we view the fully relaxed configuration in Fig. 6(b) as the global energy minimum for the zz-3-c step. Note that, in the zz-3-c case, the chemical potential $\mu$ of a H atom for a configuration is not equal to the corresponding $E_{\text{bind}}$, because the total number of H atoms at the zz-3 step is greater than one [33].

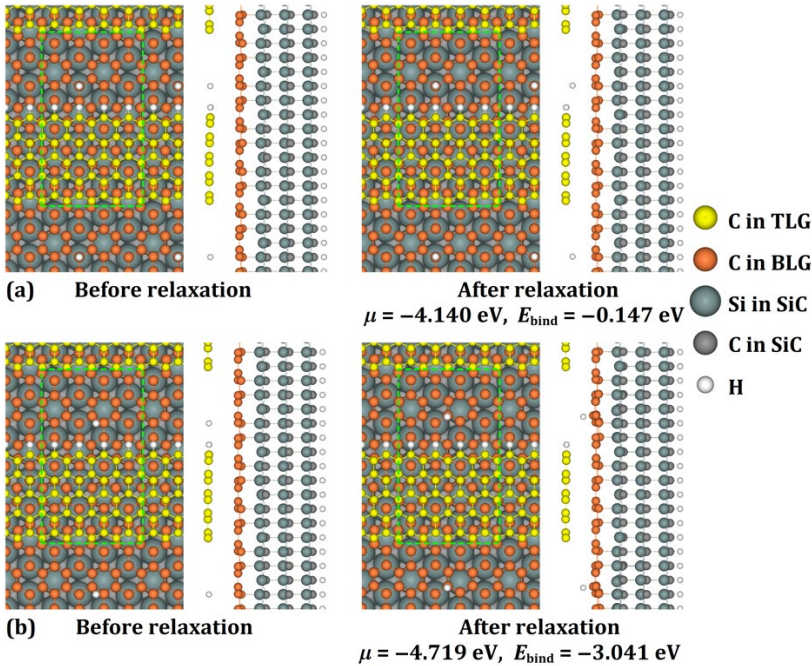

**FIG. 9.** Top and side views of structures for one H atom at zz-3-c step edges in the $2\sqrt{3}a_{\text{SiC}} \times 6a_{\text{SiC}}$ supercell (green dashed frame) used in our DFT calculations. The zz-3-c step is initially obtained by relaxing one H chain to saturate the zz-3 step. Note that the initial positions of the H atom in (a) and (b) are different. Other details are similar to the caption in Fig. 4.

### E. Overview of interactions between H and TLG steps

The chemical potential $\mu$ for systems with a single H atom, or more generally the binding energy, $E_{\text{bind}}$, reflects the average interaction strength between one H atom and its surroundings. Lower (higher) $\mu$ or $E_{\text{bind}}$ corresponds to a stronger (weaker) interaction of the H atom with other atoms. From Figs. 4 and 5 or Table S3 [33], the $\mu$ value for zz-3 is only about 0.3 eV lower than that for zz-1. In contrast, $\mu$ of a Dy atom for zz-3 is about 1.5 eV higher than that for zz-1 [32]. This indicates that the exposed BLG strip width has a much smaller effect on the covalent-like C-H bonding than on metal-C (like Dy-C) interactions. Also, with increasing strip width of exposed BLG, the interaction of H with the steps becomes stronger, while the interaction of Dy with the steps becomes weaker. From Figs. 5 and 6 or Table S3 [33], the $\mu$ value for zz-3-c is about 0.7 eV higher than that for zz-3. Thus, once a step is saturated by H (e.g., by a H chain like zz-3-c in Fig. 6), then the average interaction is significantly weakened. In summary, the order of the interaction strength from strong to weak for the zigzag steps is zz-3 > zz-1 > zz-3-c. For armchair steps, by comparing ac-1 in Fig. 7 with ac-3 in Fig. 8 or Table S3 [33], the interaction does not significantly change, indicating that increasing the strip width of exposed BLG do not affect the short-range covalent-like C-H bonding.

From results in Fig. 4 to Fig. 9 or Table S3 [33], the order from lower to higher chemical potential for H (or from stronger to weaker average interaction with H) of the three types of steps is: zigzag then armchair then zz57. As an aside, this order is the same as that for Cu at graphite steps [25], as well as Dy at TLG steps of the graphene-SiC system [32].

Here we mention that to determine the global minimum of energy for H at a specific step edge, one should in principle relax configurations as many as possible. However, it is often the case that the global minimum can be obtained by judiciously relaxing limited number of initial configurations. From our selected tests, we find that more initial positions at a step edge for the guest H atom than those in Secs. IV A, IV B, and IV C do not result in new configurations (e.g., during relaxation, the guest H atom moves to the positions in Figs. 4–9). Thus, we believe that our selection of two or three configurations with the H atom at a 1D-like step edge in one unit cell is sufficient.

## V. ENERGY LANDSCAPE AND KINETICS OF INTERCALATION VIA TLG STEPS VERSUS DIRECTLY

In this section, we discuss the kinetics of a H atom intercalating into the gallery via a TLG step versus via direct penetration through the TLG.

### A. Energy Landscape

Availability of $E_{bind}$ values for a H atom on the TLG, at various TLG steps, in the BT gallery, and also in the BS gallery, allows generation of a binding energy landscape. Each black line in Fig. 10(a) corresponds to the lowest $E_{bind}$ value from our DFT calculations in Sec. III or Sec. IV for a H atom at each of the above different positions or step edges. As already reported in Sec. III, the thermodynamic favorability of a H atom on TLG, in the BT gallery, and in the BS gallery has the order of BT > BS > top. As shown in Fig. 10(a), all $E_{bind}$ values for various steps are lower than the value in BT gallery. Therefore, the favorability order after considering these steps is updated as zz-3 > zz-1 > ac-3 > ac-1 > zz57-3 > zz-3-c > BT > BS > top, as indicated by the lines with the $E_{bind}$ values from lower to higher in Fig. 10(a). Because $E_{bind}$ of a H atom on TLG is about 1–3 eV higher than that at various step edges or about 1 eV higher than that in the BT gallery [Fig. 10(a)], there is a thermodynamic driving force for the H atom to move from the top terrace to a step edge or to the BT gallery.

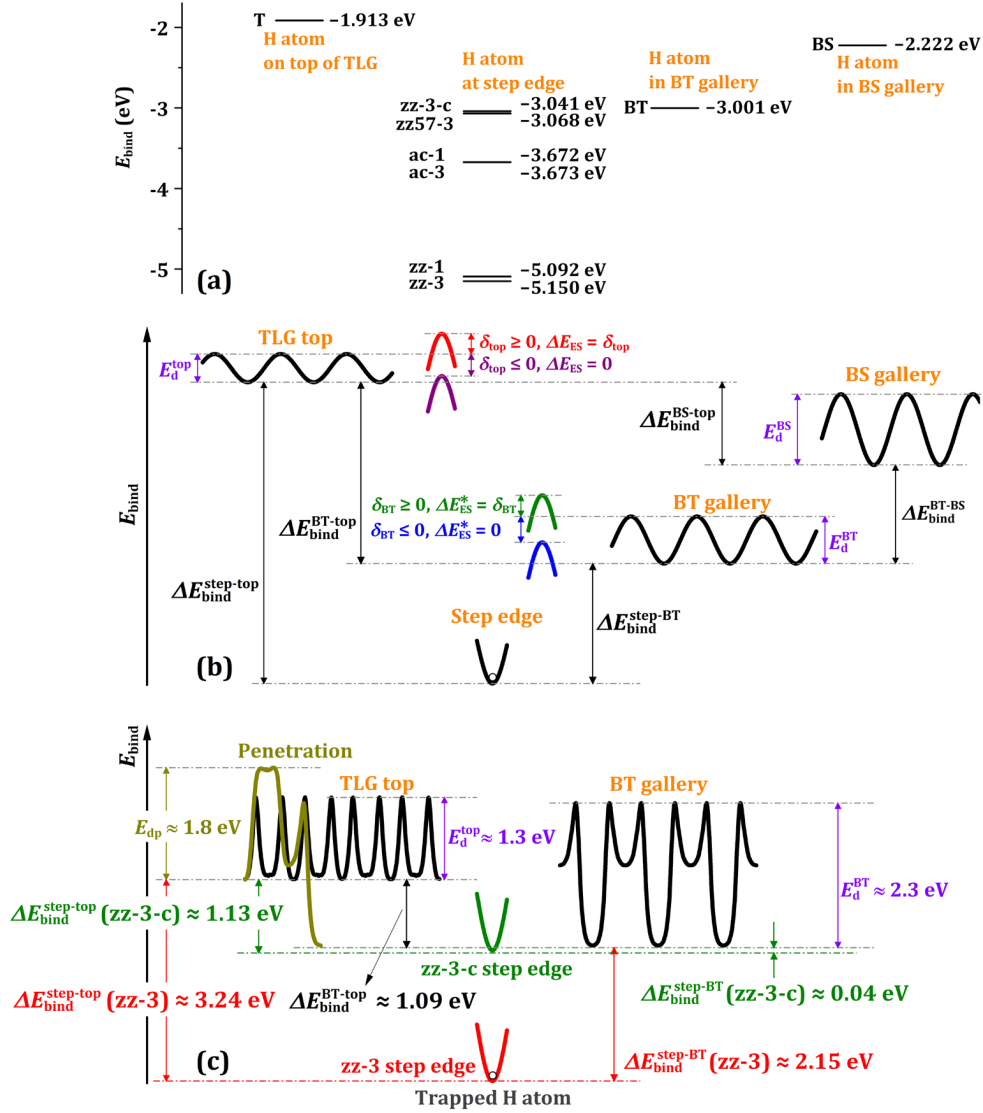

**FIG. 10.** (a) DFT binding energies ($E_{bind}$) of one H atom on TLG, at various step edges, in BT gallery, and in BS gallery. (b) Generic schematic of binding energy landscape for any guest atom (such as H) on TLG, at a step, in BT gallery, and in BS gallery. Various diffusion barriers and the binding energy differences are indicated. (c) H-specific MEPs from our CINEB calculations for one H atom on TLG and in BT gallery, as well as direct penetrating TLG. The zz-3 step edge trapping the H atom is indicated schematically by the red curve as the lower limit of $E_{bind}$ for steps in (a). The zz-3-c step edge is indicated schematically by the green curve as the upper limit of $E_{bind}$ for steps in (a). The key binding energy differences are also indicated.

To conveniently characterize the energetics (and thus kinetics) of intercalation, we first present a generic schematic of the energy landscape in Fig. 10(b) for a general guest atom on TLG, at a step, in BT gallery, and in BS gallery. Such a schematic reflects the diffusion barriers for the guest atom on TLG and in the galleries, as well as possible additional Ehrlich–Schwoebel (ES)-type barriers. For crossing a step on a surface from upper to lower terrace, the atom generally needs to overcome an extra barrier, i.e., the ES barrier $\Delta E_{ES}$, relative to the terrace diffusion barrier $E_d^{top}$. In principle, one can perform DFT-CINEB calculations to obtain $\Delta E_{ES}$, but such calculations are highly demanding. Instead, we make a rough estimate by defining an ES barrier parameter $\delta_{top}$, which is the energy difference between the transition states of the guest atom at the step edge and at a TLG terrace position far from the step. If $\delta_{top} \geq 0$, $\Delta E_{ES} = \delta_{top}$, but if $\delta_{top} \leq 0$, $\Delta E_{ES} = 0$ [53], as indicated by the red and purple curves in Fig. 10(b).

In Fig. 10(c), we also show the schematic energy landscape specifically for a guest H atom in the graphene-SiC system. This landscape is based upon the MEPs obtained in Sec. III B from our CINEB calculations for a H atom on TLG and in BT gallery, as well as direct penetrating TLG. Also, this energy landscape specifically shows the case of trapping the H atom at the zz-3 step edge. For the H atom crossing the step from upper to lower terrace, the total barrier $\Delta_{step} = E_d^{top} + \Delta E_{ES}$ should be close to $E_d^{top} = 1.31$ eV (obtained in Sec. III B) for an expected small $\Delta E_{ES}$. Thus, $\Delta_{step}$ is smaller than the penetration barrier $E_{dp} \approx 1.8$ eV (see Sec. III B). Therefore, a H atom can potentially diffuse on TLG, cross the TLG step edge, and then attach to a step site on BLG.

Once a guest atom (like H) crosses the TLG step edge and attaches to a step site on BLG, it can detach from the step and then either return to the TLG terrace or intercalate into the BT galley. The detachment from the step site back to the TLG terrace needs to overcome an energy barrier $\Delta_{back} = \Delta E_{bind}^{step-top} + E_d^{top} + \Delta E_{ES}$, where $\Delta E_{bind}^{step-top}$ is the binding energy difference from a step edge to TLG terrace [see Fig. 10(b)]. Given $\Delta E_{bind}^{step-top}$ in the range of 1.13–3.24 eV for various step types [see Figs. 10(a) and 10(c)] and neglecting $\Delta E_{ES}$, one has that $\Delta_{back} = \Delta E_{bind}^{step-top} + E_d^{top} + \Delta E_{ES}$ is in a range of about 2.4–4.5 eV.

Detachment from a step site into the BT gallery has an energy barrier $\Delta_{in} = \Delta E_{bind}^{step-BT} + E_d^{BT} + \Delta E_{ES}^*$, where $\Delta E_{bind}^{step-BT}$ is the binding energy difference from the step site to the BT gallery, $E_d^{BT} = 2.29$ eV (obtained in Sec. III B for a guest H atom) is the diffusion barrier in BT gallery, and $\Delta E_{ES}^* \geq 0$ is an ES-like intercalation barrier parameter. $\Delta E_{ES}^* = \delta_{BT}$ for $\delta_{BT} \geq 0$, but $\Delta E_{ES}^* = 0$, for $\delta_{BT} \leq 0$, corresponding to the green or blue curves in Fig. 10(b), respectively. Given $\Delta E_{bind}^{step-BT}$ in a range of about 0.04–2.15 eV for various step types [see Figs. 10(a) and 10(c)] and neglecting $\Delta E_{ES}^*$, one has that $\Delta_{in} = \Delta E_{bind}^{step-BT} + E_d^{BT} + \Delta E_{ES}^*$ is in a range of about 2.3–4.4 eV.

## B. Kinetics of Intercalation

A basic analysis of kinetics compares the rates for relevant thermally activated processes, where these rates have an Arrhenius form $r = \nu e^{-\beta E_{act}}$, where $E_{act}$ is the relevant activation barrier, $\nu \approx 10^{13}$/s is the attempt frequency, and $\beta = 1/(k_B T)$ is the inverse temperature with temperature $T$ and Boltzmann constant $k_B$. From experiments by Lin et al. [16], the temperature for H intercalation into graphene on 6H-SiC(0001) substrate is higher than 970 K. From experiments by Kunc et al. [17], the H intercalation temperature for graphene on 4*H*-SiC(0001) substrate is 1060 to 1780 K. At the lower-limit temperature $T \approx 1000$ K in these H intercalation experiments, the rate for direct penetration from the TLG terrace into the BT gallery is given by rate $r_{dp} = ve^{-\beta E_{dp}} \approx 8480/s$ if the penetration barrier $E_{dp} \approx 1.8$ eV from Model 2 is used. In contrast, the rates for the zz-3-c step detachment at 1000 K, $r_{back} = ve^{-\beta \Delta_{back}} \approx 8/s$ with the lower limit about 2.4 eV of $\Delta_{back}$ (see Sec. V A) and $r_{in} = ve^{-\beta \Delta_{in}} \approx 26/s$ with the lower limit about 2.3 eV of $\Delta_{in}$ (see Sec. V A again), are far below $r_{dp}$. Thus, it might appear that direct penetration is the dominant intercalation pathway. However, there are some caveats to this conclusion.

One significant caveat is the possibility that the actual value of $E_{dp}$ could be higher. This would both decrease the rate for direct penetration and impact the likelihood for the H adatom to reach a step or defect before undergoing direct penetration, as discussed below. The value of $E_{dp} \approx 1.8$ eV was obtained from Model 2, in which the overlayer graphene has a tensile strain of about 8.8% in order to match the SiC(0001) substrate (see Sec. II). The presence of tensile strain (stretching the graphene sheet) might be expected to lower the barrier for penetration. A test of this proposal is provided by analysis of penetration of (unstrained) freestanding bilayer graphene where CINEB analysis produced a barrier of 4.2 eV [39]. This result is reasonably consistent with an earlier DFT-LDA estimate of 3.73 eV for unstrained freestanding single-atom-thick graphene (which decreases to 0.89 eV upon introducing 10% tensile strain) [12]. These observations suggest the strong possibility that, for a refined model of supported graphene on SiC such as Model 1 with a much less tensile strain of about 0.4%, $E_{dp}$ could be significantly higher than 1.8 eV. This would shift the competition between step-mediated and direct intercalation to favor the former. A further caveat is that, even if Model 1 with strain of 0.4% is more realistic, regions with local tensile strain larger than 0.4% are likely possible in the actual system [32, 39].

For intercalation at steps, not only should the rate be higher than for direct penetration, but it is also necessary that deposited H be able to diffuse to the step before undergoing direct penetration. The lifetime of the guest H atom staying on the TLG before direct penetration is given by $\tau_{dp} = 1/r_{dp}$, and the mean time per hop on the TLG is given by $\tau_d^{top} = 1/r_d^{top}$, where $r_d^{top} = ve^{-\beta E_d^{top}}$. Thus, the mean number of hops made by the guest atom on the TLG before direct penetration is given by $N_d^{top} = \tau_{dp}/\tau_d^{top} = r_d^{top}/r_{dp} = e^{\beta(E_{dp} - E_d^{top})}$. The corresponding root-mean-square displacement of the hopping guest atom is given by [54] $d_{rms} \approx \sqrt{N_d^{top}} a_C^*$. Using $E_{dp} \approx 1.8$ eV from Model 2, one obtains $N_d^{top} \approx 330$ at 1000 K, so that $d_{rms} \approx 18 a_C^*$, i.e., 18 lattice constants of the TLG, far below the typical distance to steps or other defects. Thus, intercalation would occur almost exclusively due to direct penetration. If $E_{dp}$ is somewhat higher, say 2.5 eV, then $r_{dp} \approx 2.5/s$ at 1000 K, and $d_{rms} \approx 1060 a_C^*$. Thus, prior to direct penetration, the atom could possibly reach steps which are separated by a few thousand lattice constants on a good sample, and more readily reach other domain boundary defects which are more common and can provide access to the gallery. However, since $r_{dp}$ is comparable to the lowest value for intercalation at steps, one expects that direct penetration still dominates. If $E_{dp}$ is even higher, say 3.0 eV, then $r_{dp} \approx 0.008/s$ at 1000 K, and $d_{rms} \approx 19200\, a_C^*$. Thus, H adatoms can readily reach steps, and potentially undergo intercalation at those locations (at least for H saturated steps).

This picture might be compared with that for intercalation of a Dy atom into the BT gallery, where the intercalation via a specific step like zz-3-c or zz57-3 is much easier than direct penetration from top to BT gallery [32]. This feature reflects the large direct penetration barrier ($E_{dp} \gtrsim 3.5$ eV) compared with the small BT-gallery diffusion barrier ($E_d^{BT} = 0.54$ eV) for a Dy atom [32].

**VI. FURTHER DISCUSSION INCLUDING BEHAVIOR FOR HIGHER H COVERAGES**

From the analysis from the Sec. V B, the competition between direct penetration and intercalation at steps is impacted not just by the relative rates of these processes, but also by the capability of deposited atoms to reach steps before direct penetration. For the lower $E_{dp}$ predicted by Model 2 with large tensile strain, the lifetime of atoms on the TLG before direct penetration is too short for them to reach steps or other defects. However, for higher values of $E_{dp}$ anticipated for refined models with lower strain, it becomes plausible for H adatoms to reach steps. Then, the extent of intercalation at steps is controlled by the detaching barrier from the steps to the BT gallery. Bare or undecorated steps constitute deep traps of diffusing H atoms for which detachment into the BT gallery is not viable. However, subsequent to H deposition, these steps should be quickly saturated with H which greatly reduces the binding energy of H at the steps, and thus the barrier for detachment into the BT gallery. For example, $\Delta E_{bind}^{step-BT}$ is reduced from about 2.15 eV for the undecorated step (zz-3) to about 0.04 eV for the decorated step (zz-3-c) [see Figs. 10(c)]. The above analysis is in the low H coverage regime, and other factors come into play at higher coverages explored in experiments which will impact both the direct penetration rates and the rate for detachment from steps.

Experimentally, the importance of domain boundaries for intercalation between graphene of different thickness has been shown in several studies. These domain boundaries are separated by steps [55, 56], as studied theoretically in the current work. For H, even antiphase boundaries were found to be entry portals of intercalation and deintercalation at about 1270 K in real time with LEEM [18]. This suggests that H can move below graphene at locations which distort the local *sp*$^2$ bonding far less than steps as discussed here. This is consistent with the current conclusion that either direct penetration or by detachment from steps are both likely pathways for H to move below.

The above analysis focused on the low H-coverage regime. Thus, we now turn to the issue of whether behavior, in particular the thermodynamic preference for population of galleries versus the top surface, is affected by higher H coverages. Figure 11 summarizes our DFT predictions for the variation of $\mu$ values with H coverages (1/16, 1/4, and 1 ML). The DFT results for 1/4 and 1 ML are provided in Appendix. For 1/16 ML (corresponding to the above results), the favorability order of H is BT > BS > top with the differences of about 0.3 to 1.1 eV in $\mu$. For 1/4 ML, the order is also BT > BS > top with the differences of about 0.5 to 1.3 eV in $\mu$. However, significantly, for 1 ML, the order becomes BS > BT > top with the differences of about 0.2 to 1.2 eV in $\mu$. Thus, the interaction between a H atom and its surroundings varies strongly with coverage, where this interaction is expected to derive mainly from electronic contributions rather than elastic contributions. Note that the switch in the intercalation preference from the BT gallery to the BS gallery for higher H coverages is consistent with experimental observations where H intercalation is shown to decouple BLG from the SiC substrate [6, 7, 10, 15]. Such experimental studies are carried out commonly at high intercalation coverages, and in many cases, it was assumed or measured to be close to 1 ML. For example, it was shown that the preferred intercalation site is the BS gallery where the BLG was converted into single layer graphene [6]. In addition, the corresponding increase in the BS spacing was 78% [13], which is another factor to favor the BS gallery as the preferred intercalation location at 1 ML (see Appendix).

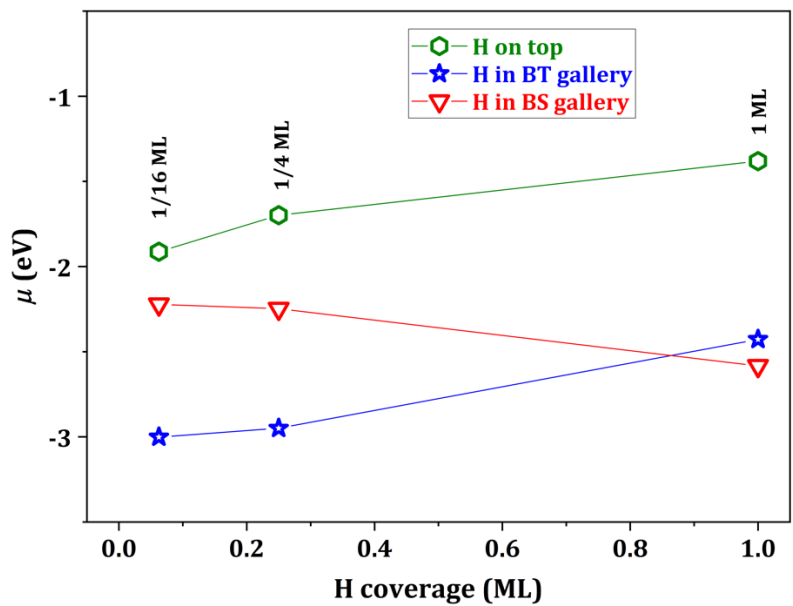

**FIG. 11.** Coverage dependence of chemical potential ($\mu$) with H on TLG, in BT gallery, and in BS gallery. DFT data are from Table S2 [33].

**VII. CONCLUSIONS**

We have performed first-principles DFT calculations for H adsorption and intercalation for a graphene-SiC system. From a chemical-potential analysis relevant for the low H-coverage regime, we find that a single H atom intercalated into BT gallery is about 1.1 eV more favorable than adsorption on TLG, and about 0.8 eV more favorable than intercalated into BS gallery. We obtain the diffusion barriers about 1.3 eV and 2.3 eV for a single H atom diffusing on TLG and in BT gallery, respectively. We also explore the competition between intercalation via direct penetration of the TLG and intercalation via detachment from steps. The competition depends critically on the barrier for direct penetration. For the lower estimate from Model 2 with large tensile strain, direct penetration dominates. However, for higher values anticipated from refined models with less strain, the competition can switch so intercalation at steps dominates provided that steps are saturated with H atoms. Finally, we note the significant observation that preference for intercalation switches from the BT gallery to the BS gallery for higher H coverages. At the H coverage of 1 ML, the interlayer spacing for the BT gallery become significantly larger so that a quasi-freestanding graphene layer is effectively obtained, consistent with experimental observations. These theoretical results and analyses provide insights for experiments of H intercalation.

**SUPPLEMENTARY MATERIAL**

See Supplementary Material for i) formulation of chemical potential, adsorption energy, binding energy, combination energy, and intercalation energy; ii) DFT data for the configurations in Figs. 1, 2, 12 and 13; iii) DFT data for the configurations in Figs. 4, 5, 6, 7, 8, and 9; iv) the projected density of states (PDOS) for analyzing the spin polarization of the different atoms in various configurations.

**ACKNOWLEDGMENT**

Work by Y. H. and M. C. T was supported mainly by the U. S. Department of Energy (DOE), Office of Science, Basic Energy Sciences (BES), Materials Sciences and Engineering Division. Work by J. W. E. was supported the USDOE BES Division of Chemical Sciences, Geosciences, and Biosciences through the Computational and Theoretical Chemistry program. Research was performed at the Ames Laboratory, which is operated by Iowa State University under contract No. DE-AC02-07CH11358. DFT calculations were mainly performed with a grant of computer time at the National Energy Research Scientific Computing Centre (NERSC). NERSC is a DOE Office of Science User Facility supported by the Office of Science of the U. S. DOE under Contract No. DE-AC02-05CH11231. The calculations also partly used the Extreme Science and Engineering Discovery Environment (XSEDE), which is supported by National Science Foundation under Grant No. ACI-1548562.

**APPENDIX: ADSORPTION AND INTERCALATION OF H WITH HIGHER COVERAGES**

In experiments [6, 10, 13, 15], the coverage for H intercalation is often much higher than 1/16 ML described in Sec. III. Thus, we also choose two higher coverages of 1/4 and 1 ML to assess the energetics of H adsorption and intercalation in this section. In our DFT calculations for these higher coverages, we use the same supercell and $k$ mesh as those in Sec. III, and therefore the coverages 1/4 ML and 1 ML correspond to 4 and 16 H atoms in the supercell, respectively. We only consider the perfect graphene, i.e., without any steps, for this section.

### A. 1/4 ML H

For H adsorption on TLG, intercalation in BT gallery, or intercalation in BS gallery, we always initially select one uniform configuration. After full relaxation, we obtain the configurations with the chemical potentials $\mu$ shown in Fig. 12. We find that within a supercell the four H atoms on TLG in Fig. 12(a) or in BS gallery in Fig. 12(c) have negligible lateral relaxations relative to the initial configuration, but the four H atoms in BT gallery in Fig. 12(b) tend to clustering instead of the initial uniform H layer for this coverage.

The $\mu$ values of three fully relaxed configurations in Figs. 12(a), 12(a), and 12(c) are −1.698 eV, −2.950 eV, and −2.247 eV, respectively. To judge the energetic favorability of H on TLG, in BT gallery, or in BS gallery, in principle, one needs to relax a large number of configurations [32, 26]. In this work, we only select one configuration for each case, and thus we can conservatively say that the 1/4-ML H is likely energetically most favorable in BT gallery than on TLG and in BS gallery, i.e., with an uncertain favorability order of BT > BS > top, based on the fact that the $\mu$ value in Fig. 12(b) is significantly (1.252 eV and 0.703 eV, respectively) lower than those in Figs. 12(a) and 12(c).

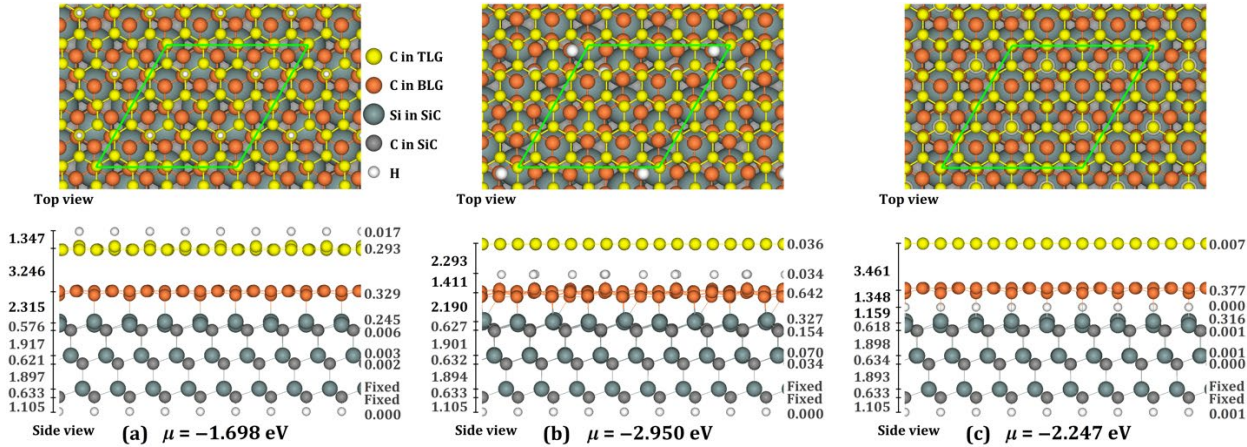

**FIG. 12.** Fully relaxed structures with a H coverage of 1/4 ML (a) and (b) on TLG, (c) in BT gallery, and (d) in BS gallery from our DFT calculations. The four H atoms in the $2\sqrt{3}a_{SiC} \times 2\sqrt{3}a_{SiC}$ supercell (green frame in top view) are always initially set to be uniformly distributed. The left of side view shows the interlayer spacings (in Å) and the right shows the corrugations (in Å) for each single-atom-thick layer (for definitions of interlayer spacings and corrugations, see the caption of Fig. 1).

## B. 1 ML H

Similarly, for 1 ML H (i.e., 16 H atoms per cell), we also initially select one uniform configuration for adsorption on TLG, intercalation in BT gallery, or intercalation in BS gallery. After full relaxation, we obtain the configurations with the chemical potentials $\mu$ shown in Fig. 13. We find that within a supercell the 16 H atoms on TLG in Fig. 13(a) have negligible lateral relaxations relative to the initial configuration, but the 16 H atoms in BT gallery in Fig. 13(b) or in BS gallery in Fig. 13(c) have noticeable relaxations, especially along the direction vertical to the layers. In Fig. 13(b), both the H layer and BLG (with larger corrugations 1.556 and 1.226 Å, respectively) have larger relaxations, where some H atoms can intermix with BLG. In Fig. 13(c), both the H layer and the terminal Si layer (with larger corrugations 0.740 and 0.571 Å, respectively) have larger relaxations. These larger relaxations plausibly reflect the preference of H tending to bond with Si atoms at this coverage. This is also consistent with the order of $\mu$ values from lower to higher: $-2.583$ eV, $-2.427$ eV, and $-1.381$ eV in Figs. 13(c), 13(b), and 13(a), respectively, i.e., the favorability order is BS > BT > top. As mentioned above, this order has uncertainty because we only select one configuration for H on TLG, in BT gallery, and in BS gallery.

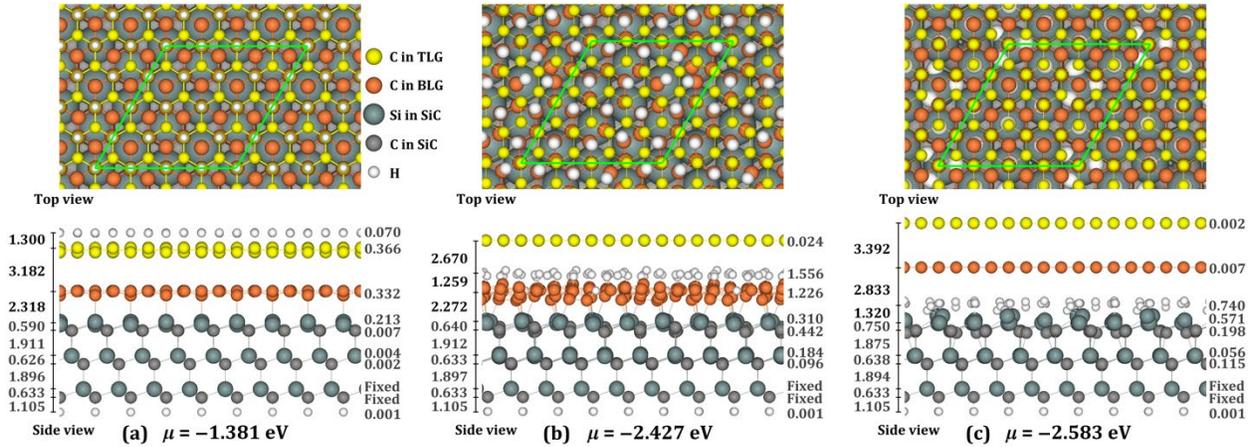

**FIG. 13.** Fully relaxed structures with a H coverage of 1 ML (a) on TLG, (b) in BT gallery, and (c) in BS gallery from our DFT calculations. The sixteen H atoms in the $2\sqrt{3}a_{SiC} \times 2\sqrt{3}a_{SiC}$ supercell (green frame in top view) are always initially set to be uniformly distributed. The left of side view shows the interlayer spacings (in Å) and the right shows the corrugations (in Å) for each single-atom-thick layer (for definitions of interlayer spacings and corrugations, see the caption of Fig. 1).

## C. Coverage dependences of chemical potentials and interlayer spacings

Figure 11 in the main text already summarizes the dependence of chemical potentials on H coverage. In Fig. 14, we summarize the coverage dependence of interlayer spacings $d_{BT}$ and $d_{BS}$. The data for all interlayer spacings and corrugations are shown in Figs. 2, 11, and 12 (also see Table S1 [33]). From Fig. 13(b), any spacing without H intercalated has almost no change (the change less than 6%) for any coverage relative to 0 ML (i.e., without H). After a gallery is intercalated, the spacing for a lower coverage (1/16 or 1/4 ML) still has little change (the change less than 10%) relative to 0 ML, but the spacings for 1 ML significantly increase (also see the values highlighted by the bold fonts in Table S1 [33]). $d_{BT}$ and $d_{BS}$ increase by about 16% and 78% after 1 ML H intercalates into BT and BS galleries, respectively. This feature contrasts with the intercalation of 1/16, 1/4, or 1 ML Dy [32], for which the interlayer spacings for any BT or BS gallery *without* the Dy atom intercalated have small changes less than 7%, but the interlayer spacing for any BT or BS gallery *with* the Dy atom intercalated has a large increase of about 22% at least. The intercalated or adsorbed Dy for high coverage (1 ML) become two single-atom-thick Dy layers after full relaxations (and therefore the interlayer spacing of a Dy-intercalated BT or BS gallery has an increase up to about 100% to 200%) [32], but the intercalated or adsorbed H layer is always single-atom-thick although the intercalated H layer and its adjacent BLG [Fig. 12(b)] or Si layer [Fig. 12(c)] have significant relaxations or corrugations, as already described in Sec. VI B.

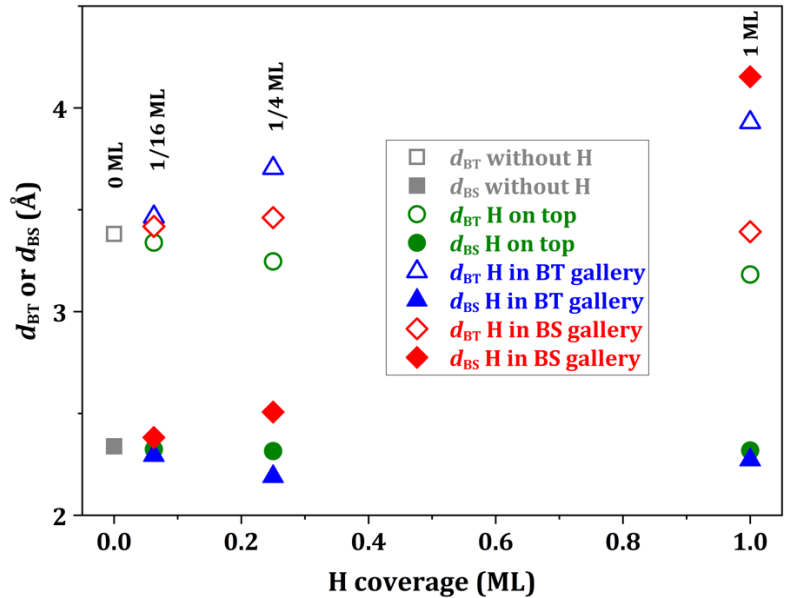

**FIG. 14.** Coverage dependence of interlayer spacings ($d_{BT}$ and $d_{BS}$) without or with H on TLG, in BT gallery, and in BS gallery. DFT data are from Table S1 [33].

Derived from an analysis of high-resolution x-ray reflectivity, Emery, et al. [13] obtain $d_{BS} = d_{H-Si} + d_{H-BL} = 4.22$ Å (where $d_{H-Si} = 1.5$ Å is the interlayer spacing between the intercalated H layer and the terminal Si layer, and $d_{H-BL} = 2.72$ Å is the interlayer spacing between the intercalated H layer and the BLG) and $d_{BT} = 3.35$ Å. This experimental result is in good agreement with our DFT values $d_{BS} = d_{H-Si} + d_{H-BL} = 1.320 + 2.833 = 4.153$ Å and $d_{BT} = 3.392$ Å for the intercalation of 1 ML H into BS gallery, as shown in Fig. 12(c). The large increasing of 78% in $d_{BS}$ relative to 0 ML (as described above) indicates that the TLG plus BLG becomes the so-called "quasi-freestanding" bilayer graphene after the intercalation of 1 ML H into the BS gallery. In other words, decoupling the bilayer graphene from the Si layer needs a higher H coverage like 1 ML in contrast to 1/4 ML. Here, it should be mentioned that our work in this paper focuses on only a single-atom-thick graphene layer on BLG [such case is commonly referred as single layer graphene (SLG) in previous literature]. In previous experiments for guest atom intercalation, the thicknesses of graphene overlayer are often thicker, e.g., bilayer (two single-atom-thick graphene layers) or trilayer (three single-atom-thick graphene layers) graphene on BLG, for which corresponding DFT calculations are needed because of the thickness dependence of the results.

We also should mention that, comparing Fig. 11 with Fig. 14, there is no noticeable relationship between interlayer spacings with intercalated H and the corresponding chemical potentials, unlike the intercalation of metal atoms like Dy [32], for which the elastic contribution can be dominant over electronic contribution due to the relatively larger ionic-like radius [29]. For H atom, the atomic radius is much smaller than metal atoms and the bonding with C in graphene is covalent-like [29], and thus the electronic contribution is dominant over elastic contribution, which is related to the interlayer spacings and corrugations.

Several studies have also demonstrated that, at the location where the metal bonds, there are large changes of at least about 25% in the interlayer spacing relative to the pristine graphene values [58, 59, 60]. Such changes cost elastic energy which is compensated by the bonding energy gain of atoms between the graphene sheets. Only the spacing of the metal location expands and the spacings at all other locations are unchanged. The expansion is relevant in other experiments, e.g., Ge intercalated *p*-type phase under TLG and the coverage of 2-ML Ge requires interlayer expansion to accommodate the Ge bilayer [61]. Interlayer expansion is also seen in our DFT studies of Pb [62] and Gd [63] intercalation in the graphene-SiC systems.

**DATA AVAILABILITY**

The data that supports the findings of this study are available within the article and its supplementary material.

# Supplementary Material

for "**Thermodynamics and kinetics of H adsorption and intercalation for graphene on 6*H*-SiC(0001) surface from first-principles calculations**"

Yong Han*, James W. Evans, and Michael C. Tringides

Ames Laboratory, U. S. Department of Energy, Ames, Iowa 50011, USA
Department of Physics and Astronomy, Iowa State University, Ames, Iowa 50011, USA

* y27h@ameslab.gov

## Contents

**S1. Formulation of chemical potential, adsorption energy, binding energy, combination energy, and intercalation energy**

**S2. DFT data of interlayer spacings and corrugations**

**S3. DFT data for the configurations in Figs. 1, 2, 12 and 13**

**S4. DFT data for the configurations in Figs. 4, 5, 6, 7, 8, and 9**

**S5. Projected density of states (PDOS) for analyzing the spin polarization of the different atoms in various configurations**

## S1. Formulation of chemical potential, adsorption energy, binding energy, combination energy, and intercalation energy

To assess the relative stability and the average strength of atom interactions of various structures involving H binding with the graphene-SiC system, particularly at step edges, we define a chemical potential of H as [25, 22]

$$\mu = \frac{E_{\text{tot}} - E_{\text{cln}}}{n} - E_{\text{H}}, \tag{S1}$$

where $E_{\text{tot}}$ is the total energy of the H structure plus graphene and SiC, $E_{\text{cln}}$ is the energy of the fully relaxed clean graphene-SiC system (without guest H) with or without steps, $n$ is the total number of guest H atoms in a supercell, and $E_{\text{H}}$ is the energy of one isolated H atom in vacuum. The chemical potential $\mu$ in Eq. (S1) accounts for the average strength of one H atom interacting with its surroundings. A lower (higher) $\mu$ value corresponds to a configuration with a stronger (weaker) interaction of a H atom with other atoms on average. For one adatom ($n = 1$) adsorbed on TLG of the graphene-SiC system, $\mu$ reduces to the conventional adsorption energy

$$E_{\text{ads}} = E_{\text{tot}} - E_{\text{cln}} - E_{\text{H}}. \tag{S2}$$

To provide additional insight into the energy landscape of one specific H atom bound to different sites on TLG ribbon, at a TLG step edge (with or without a H chain), and in BT gallery, we also define a binding energy for this H atom as [25]

$$E_{\text{bind}} = E_{\text{tot}} - E_{\text{cln}}^* - E_{\text{H}}, \tag{S3}$$

where $E_{\text{cln}}^*$ is the energy of the fully relaxed graphene-SiC system with or without a pre-decorated H chain. Lower $E_{\text{bind}}$ indicates stronger binding between the H atom and other atoms. For the graphene-SiC system not pre-decorated by a H chain, $E_{\text{cln}}^*$ reduces to $E_{\text{cln}}$ and then Eq. (S3) reduces to the case of $n = 1$ in Eq. (S1). In this case, $E_{\text{bind}} = \mu$.

Note that the energy reference in Eqs. (S1), (S2), and (S3) is $E_{\text{H}}$. Here we also define a combination energy [29]

$$E_{\text{comb}} = E_{\text{tot}} - E_{\text{cln}} - \sigma_{\text{H}}, \tag{S4}$$

where $\sigma_{\text{H}}$ is half of the energy of a gas-phase H$_2$ molecule, instead of $E_{\text{H}}$. Then, $E_{\text{comb}}$ quantifies the energy required for a H atom to become combined with the graphene-SiC system (e.g., adsorbed on TLG, bound at a step edge, or intercalated into a gallery), starting from a gas-phase H$_2$ molecule. Note that the difference $E_{\text{comb}} - \mu = E_{\text{H}} - \sigma_{\text{H}}$ is equal to half of the bond strength of the H$_2$ molecule. For the case of intercalation, $E_{\text{comb}}$ is often called the intercalation energy $E_{\text{int}}$ [38]. From Eq. (S4), $E_{\text{comb}} > 0$ indicates that the combination of a H atom with the graphene-SiC system from the H$_2$ molecule in gas phase is endothermic, while $E_{\text{comb}} < 0$ indicates that the combination is exothermic.

## S2. DFT data of interlayer spacings and corrugations

**TABLE S1.** Interlayer spacings and corrugations from DFT calculations for four coverages of H. For the coverage of 0 ML (i.e., clean graphene-SiC system without guest H), the data are previously obtained from Model 1 and Model 2 [32], as well as PBE-TS calculations from Sforzini, et al. [15]. For the coverages of 1/16, as well as 1/4 and 1 ML, the data are from Model 2. The coverage of 1/16 ML approximates a single isolated H atom. T, BT, and BS stand for the guest H atom on TLG, in BT gallery, and in BS gallery, respectively, as shown in Figs. 1, 11, and 12. The bold fonts highlight the interlayer spacings of intercalated galleries.

| Coverage | 0 ML [32] | | | 1/16 ML | | | 1/4 ML | | | 1 ML | | |
|---|---|---|---|---|---|---|---|---|---|---|---|---|
| | Model 1 | Model 2 | [15] | T | BT | BS | T | BT | BS | T | BT | BS |
| $d_{BT}$ | 3.472 | 3.381 | 3.40 | 3.339 | **3.465** | 3.418 | 3.246 | **3.704** | 3.461 | 3.182 | **3.929** | 3.392 |
| $d_{BS}$ | 2.525 | 2.339 | 2.36 | 2.324 | 2.294 | **2.382** | 2.315 | 2.190 | **2.507** | 2.318 | 2.272 | **4.153** |
| $d_{12}$ | 0.607 | 0.564 | 0.55 | 0.577 | 0.588 | 0.577 | 0.576 | 0.627 | 0.618 | 0.590 | 0.640 | 0.750 |
| $d_{23}$ | 1.906 | 1.917 | 1.92 | 1.915 | 1.913 | 1.914 | 1.917 | 1.901 | 1.898 | 1.911 | 1.912 | 1.875 |
| $d_{34}$ | 0.592 | 0.622 | 0.61 | 0.624 | 0.626 | 0.626 | 0.621 | 0.632 | 0.634 | 0.626 | 0.633 | 0.638 |
| $d_{45}$ | | 1.896 | 1.90 | 1.897 | 1.896 | 1.897 | 1.897 | 1.894 | 1.893 | 1.896 | 1.897 | 1.894 |
| $d_{56}$ | | 0.633 | 0.62 | 0.633 | 0.633 | 0.633 | 0.633 | 0.633 | 0.633 | 0.633 | 0.633 | 0.633 |
| $d_{C-H}$ | 1.108 | 1.105 | | 1.105 | 1.105 | 1.105 | 1.105 | 1.105 | 1.105 | 1.105 | 1.105 | 1.105 |
| $c_{TL}$ | 0.694 | 0.008 | 0.45 | 0.303 | 0.025 | 0.023 | 0.293 | 0.036 | 0.007 | 0.366 | 0.024 | 0.002 |
| $c_{BL}$ | 1.227 | 0.320 | 0.86 | 0.333 | 0.663 | 0.430 | 0.329 | 0.642 | 0.377 | 0.332 | 1.226 | 0.007 |
| $c_1$ | 0.334 | 0.306 | 0.78 | 0.255 | 0.317 | 0.342 | 0.245 | 0.327 | 0.316 | 0.213 | 0.310 | 0.571 |
| $c_2$ | 0.203 | 0.001 | 0.30 | 0.003 | 0.151 | 0.077 | 0.006 | 0.154 | 0.001 | 0.007 | 0.442 | 0.198 |
| $c_3$ | 0.110 | 0.004 | 0.21 | 0.002 | 0.070 | 0.040 | 0.003 | 0.070 | 0.001 | 0.004 | 0.184 | 0.056 |
| $c_4$ | 0.000 | 0.000 | 0.14 | 0.002 | 0.033 | 0.017 | 0.002 | 0.034 | 0.000 | 0.002 | 0.096 | 0.115 |
| $c_5$ | | 0.000 | 0.08 | 0.000 | 0.000 | 0.000 | 0.000 | 0.000 | 0.000 | 0.000 | 0.000 | 0.000 |
| $c_6$ | | 0.000 | 0.05 | 0.000 | 0.000 | 0.000 | 0.000 | 0.000 | 0.000 | 0.000 | 0.000 | 0.000 |
| $c_H$ | 0.009 | 0.000 | | 0.000 | 0.001 | 0.001 | 0.000 | 0.000 | 0.001 | 0.001 | 0.001 | 0.001 |

## S3. DFT data for the configurations in Figs. 1, 2, 12 and 13

**TABLE S2.** DFT data for chemical potential $\mu$, adsorption energy $E_{ads}$, binding energy $E_{bind}$, combination energy $E_{comb}$, and magnetic moment $m$ (in units of Bohr magneton $\mu_B$) for the configurations in Figs. 1, 2, 12, and 13. All energies or chemical potentials are in units of eV.

| Coverage | Configuration | $\mu$ | $E_{ads}$ | $E_{bind}$ | $E_{comb}$ | $m$ |
|---|---|---|---|---|---|---|
| 1/16 ML | a1, Fig. 1 | -1.805 | -1.805 | -1.805 | 0.689 | 2.9944 |
| | a2, Fig. 1 | -1.912 | -1.912 | -1.912 | 0.582 | 2.9689 |
| | a3, Fig. 1 | -1.913 | -1.913 | -1.913 | 0.582 | 2.9696 |
| | a4, Figs. 1, 2a | -1.913 | -1.913 | -1.913 | 0.581 | 2.9682 |
| | a5, Fig. 1 | -0.091 | -0.091 | -0.091 | 2.403 | 4.8710 |
| | a6, Fig. 1 | -1.912 | -1.912 | -1.912 | 0.582 | 2.9693 |
| | a7, Fig. 1 | -1.912 | -1.912 | -1.912 | 0.582 | 2.9680 |
| | a8, Fig. 1 | -1.913 | -1.913 | -1.913 | 0.582 | 2.9749 |
| | a9, Fig. 1 | -1.886 | -1.886 | -1.886 | 0.608 | 2.9995 |
| | b1, Fig. 1 | -1.723 | | -1.723 | 0.771 | 3.0000 |
| | b2, Fig. 1 | -3.001 | | -3.001 | -0.507 | 2.3264 |
| | b3, Figs. 1, 2b | -3.001 | | -3.001 | -0.507 | 2.3277 |
| | b4, Fig. 1 | -3.001 | | -3.001 | -0.507 | 2.3264 |
| | b5, Fig. 1 | 0.336 | | 0.336 | 2.830 | 4.8877 |
| | b6, Fig. 1 | -3.001 | | -3.001 | -0.507 | 2.3289 |
| | b7, Fig. 1 | -3.000 | | -3.000 | -0.506 | 2.3365 |
| | b8, Fig. 1 | -3.001 | | -3.001 | -0.507 | 2.3262 |
| | b9, Fig. 1 | -1.712 | | -1.712 | 0.782 | 3.0000 |
| | c1, Fig. 1 | -0.846 | | -0.846 | 1.648 | 2.8889 |
| | c2, Fig. 1 | -2.222 | | -2.222 | 0.272 | 0.5440 |
| | c3, Fig. 1 | -2.222 | | -2.222 | 0.272 | 0.8710 |
| | c4, Fig. 1 | -2.222 | | -2.222 | 0.272 | 0.5560 |
| | c5, Fig. 1 | -0.837 | | -0.837 | 1.658 | 0.8884 |
| | c6, Figs. 1, 2c | -2.222 | | -2.222 | 0.272 | 0.5567 |
| | c7, Fig. 1 | -2.222 | | -2.222 | 0.272 | 0.5555 |
| | c8, Fig. 1 | -2.206 | | -2.206 | 0.288 | 2.5001 |
| | c9, Fig. 1 | -2.158 | | -2.158 | 0.336 | 0.2681 |
| 1/4 ML | Fig. 12a | -1.698 | | | 0.796 | 2.1369 |
| | Fig. 12b | -2.950 | | | -0.455 | 0.1208 |
| | Fig. 12c | -2.247 | | | 0.247 | 0.0002 |
| 1 ML | Fig. 13a | -1.381 | | | 1.113 | 3.9027 |
| | Fig. 13b | -2.427 | | | 0.067 | 0.1774 |
| | Fig. 13c | -2.583 | | | -0.089 | 0.0000 |

## S4. DFT data for the configurations in Figs. 4, 5, 6, 7, 8, and 9

**TABLE S3.** DFT data for chemical potential $\mu$, adsorption energy $E_{\text{ads}}$, binding energy $E_{\text{bind}}$, combination energy $E_{\text{comb}}$, and magnetic moment $m$ (in units of Bohr magneton $\mu_B$) for the configurations in Figs. 4, 5, 6, 7, 8, and 9. All energies or chemical potentials are in units of eV.

| Step type | Configuration | $\mu$ | $E_{\text{ads}}$ | $E_{\text{bind}}$ | $E_{\text{comb}}$ | $m$ |
|---|---|---|---|---|---|---|
| zz-1 | Fig. 4a | -5.092 | -5.092 | -5.092 | -2.598 | 16.3985 |
|  | Fig. 4b | -5.090 | -5.090 | -5.090 | -2.596 | 16.3903 |
| zz-3 | Fig. 5a | -5.149 | -5.149 | -5.149 | -2.654 | 16.6582 |
|  | Fig. 5b | -5.150 | -5.150 | -5.150 | -2.656 | 16.6315 |
| ac-1 | Fig. 6a | -3.672 | -3.672 | -3.672 | -1.178 | 8.6665 |
|  | Fig. 6b | -3.666 | -3.666 | -3.666 | -1.171 | 8.6664 |
| ac-3 | Fig. 7a | -0.116 | -0.116 | -0.116 | 2.378 | 8.6666 |
|  | Fig. 7b | -3.673 | -3.673 | -3.673 | -1.178 | 8.6666 |
| zz57-3 | Fig. 8a | -0.171 | -0.171 | -0.171 | 2.323 | 12.9816 |
|  | Fig. 8b | -3.068 | -3.068 | -3.068 | -0.574 | 10.8749 |
|  | Fig. 8c | -2.906 |  | -2.906 | -0.411 | 10.9020 |
| zz-3-c | Fig. 9a | -4.140 |  | -0.147 |  | 12.4909 |
|  | Fig. 9b | -4.719 |  | -3.041 |  | 10.3171 |

## S5. Projected density of states (PDOS) for analyzing the spin polarization of the different atoms in various configurations

As listed in Tables S2 and S3, the magnetic moments of most of configurations are nonzero. To provide additional useful information for analyzing the spin polarization of the different atoms in various configurations, we selectively plot the total PDOS of each element (H, C, or Si) for the configurations listed in Table S2 and S3, as shown in Figs. S1–S11. The positive (negative) PDOS in these plots represents the spin-up (spin-down) states. An asymmetric spin-up and spin-down PDOS curve indicates the spin polarization, i.e., the system is magnetic for most of configurations, while a symmetric spin-up and spin-down PDOS curve indicates a nonmagnetic system for, e.g., Figs. 12(c) and 13(c). The $x$ and $y$ axes are always along the lateral directions, while the $z$ axis is always perpendicular to the surface and upward. The energy zero is always set to be the Fermi level $E_F$.

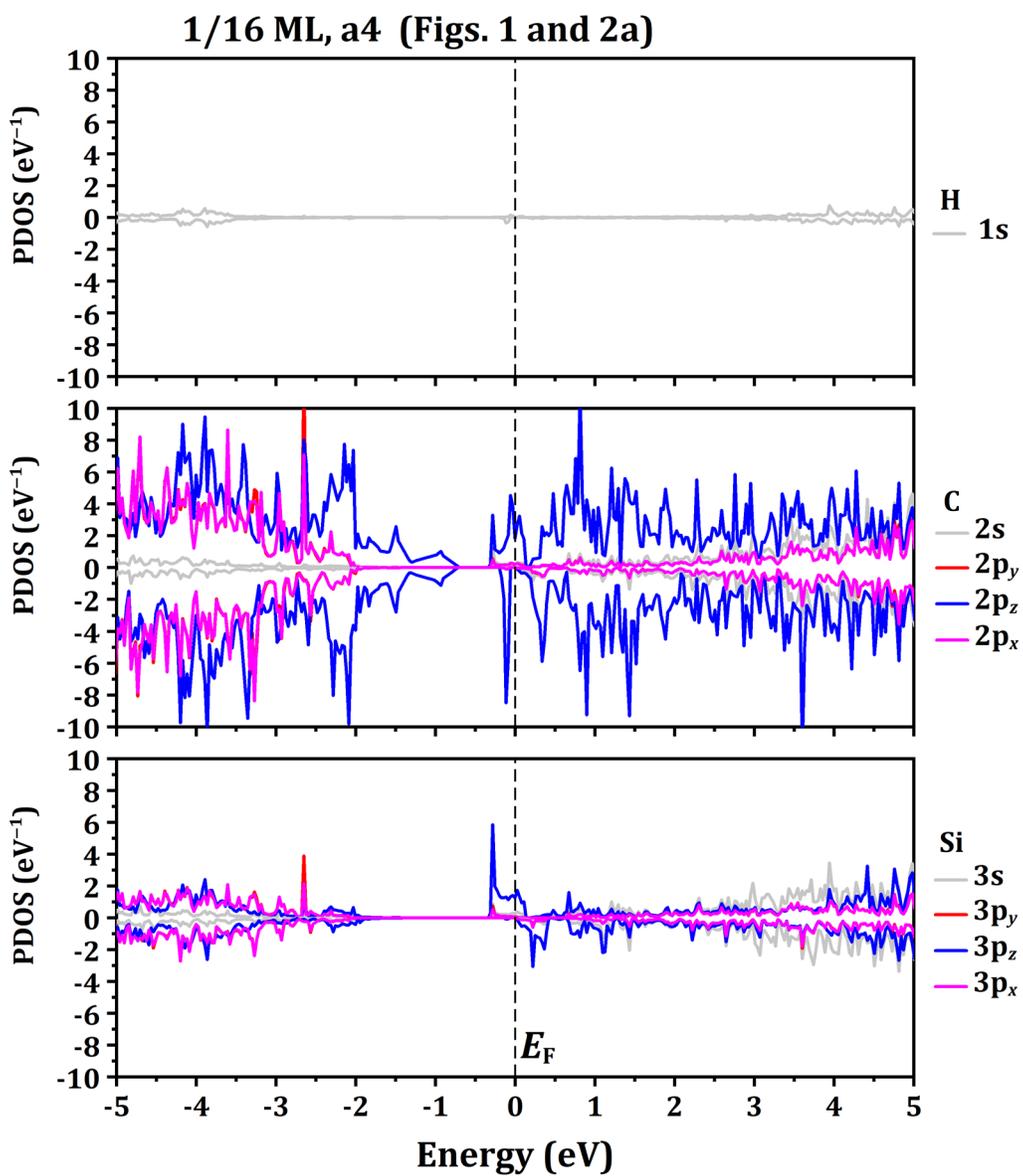

**FIG. S1.** The total PDOS of each element (H, C, or Si) for the configuration a4 in Figs. 1 and 2(a) and also listed in Table S2.

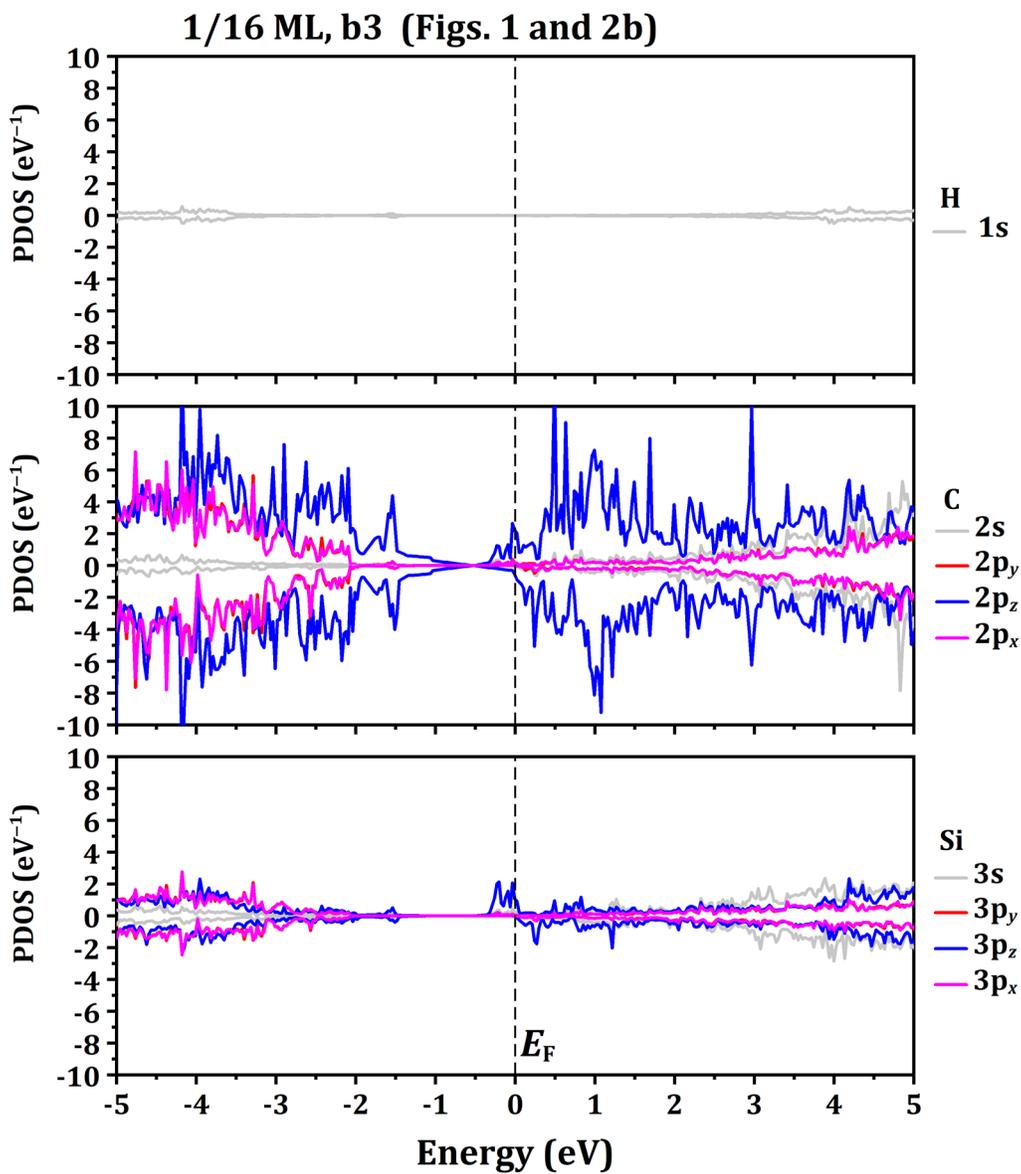

**FIG. S2.** The total PDOS of each element (H, C, or Si) for the configuration b3 in Figs. 1 and 2(b) and also listed in Table S2.

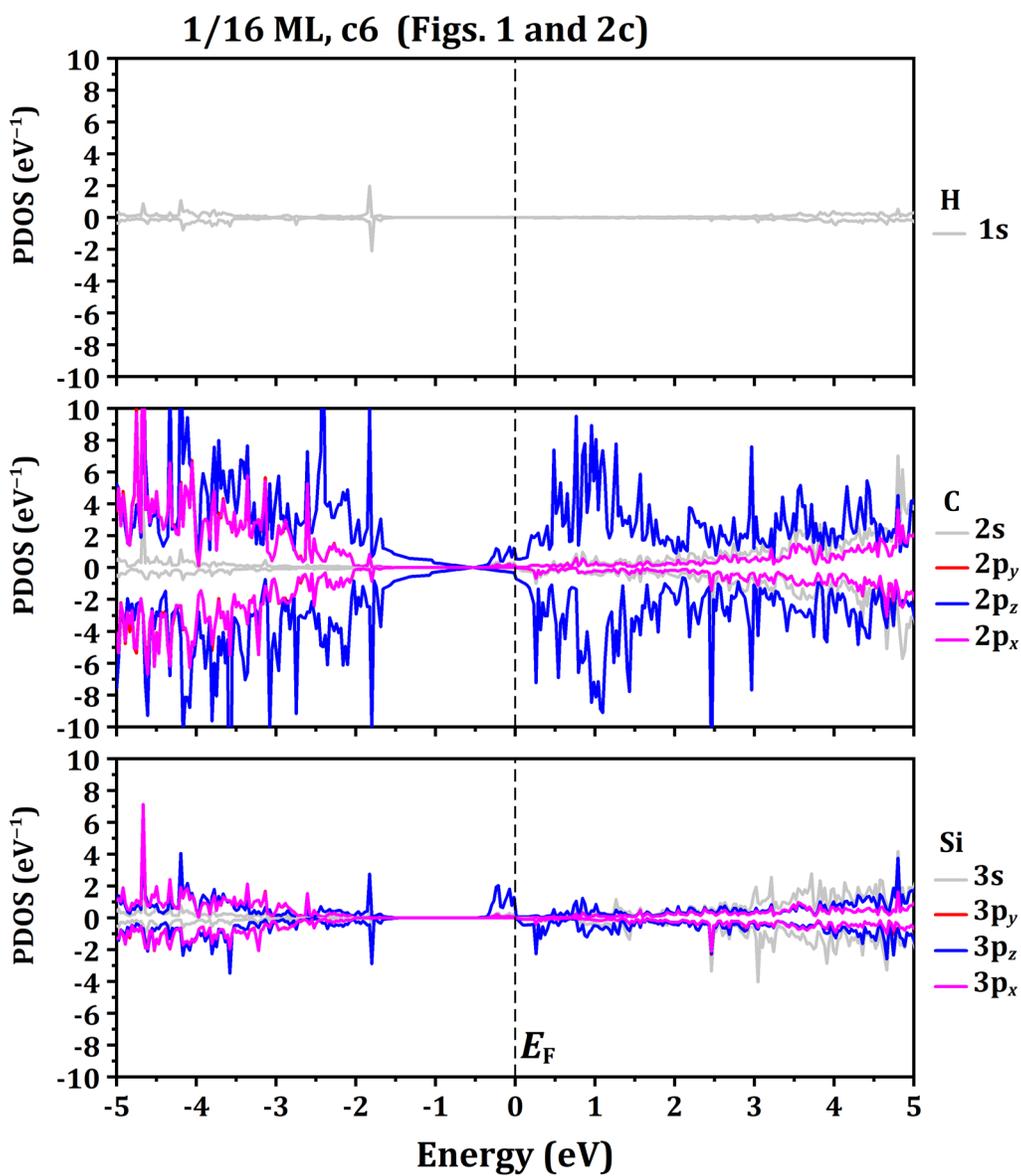

**FIG. S3.** The total PDOS of each element (H, C, or Si) for the configuration c6 in Figs. 1 and 2(c) and also listed in Table S2.

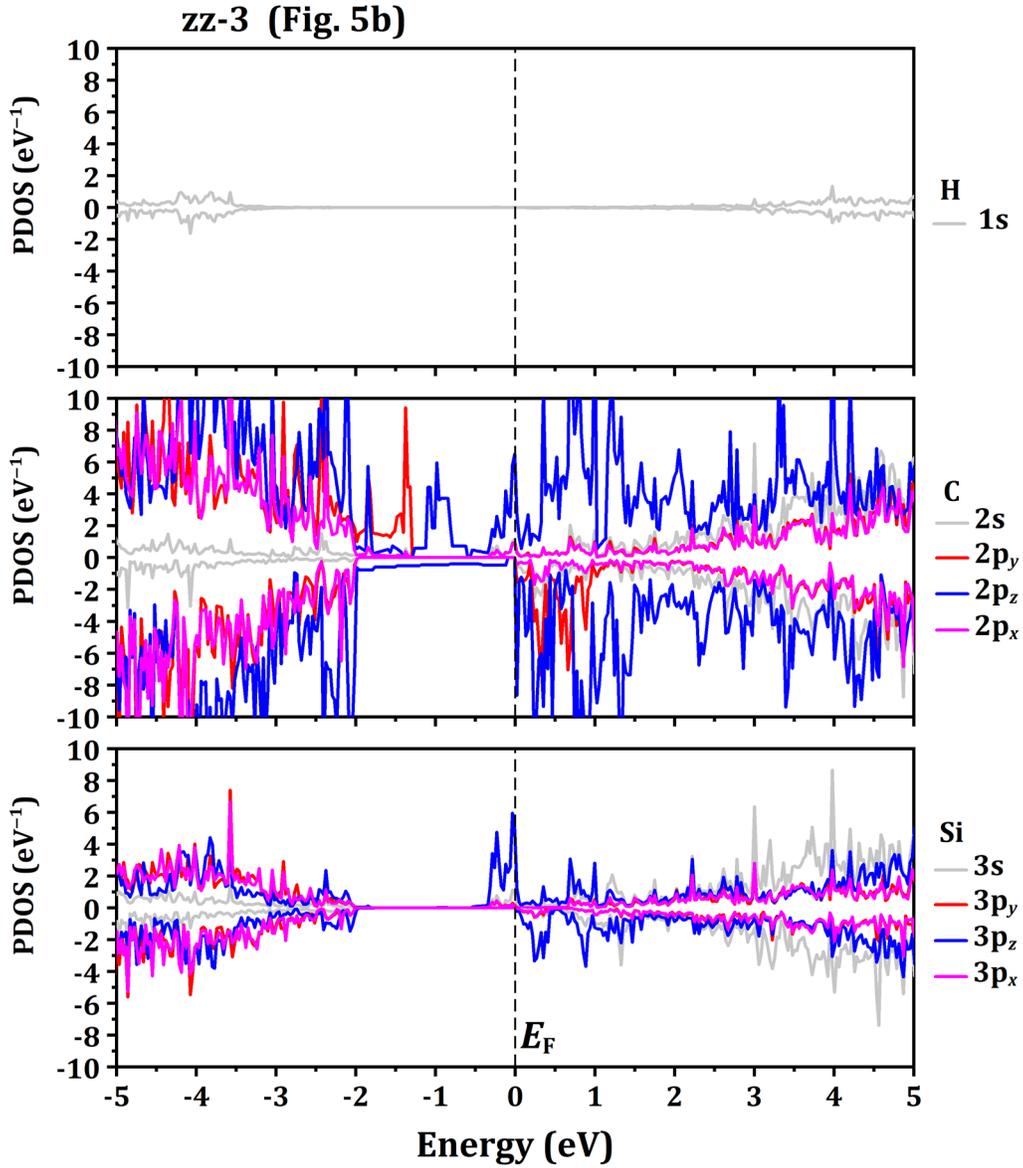

**FIG. S4.** The total PDOS of each element (H, C, or Si) for the configuration in Fig. 5(b) and also listed in Table S3.

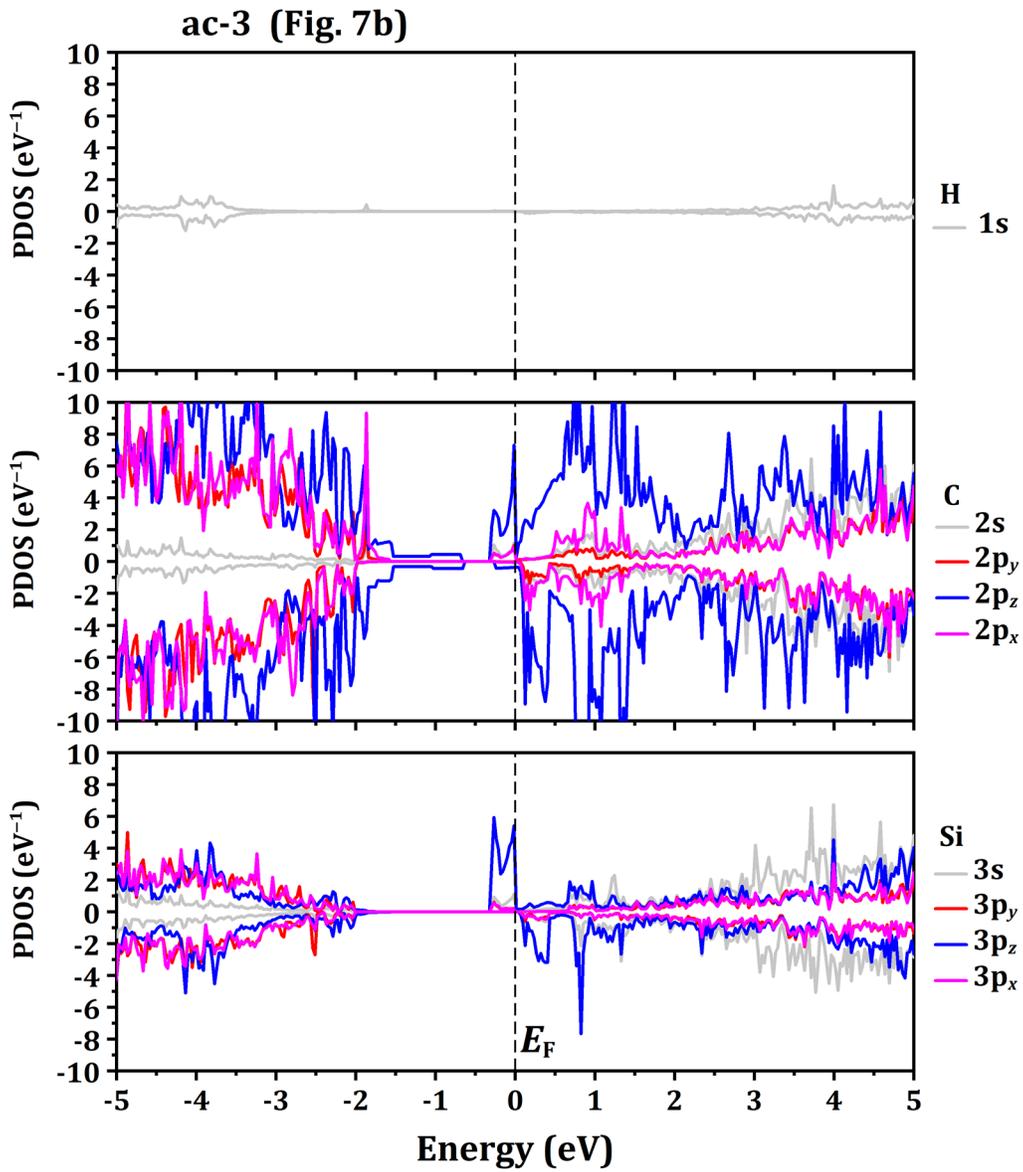

**FIG. S5.** The total PDOS of each element (H, C, or Si) for the configuration in Fig. 7(b) and also listed in Table S3.

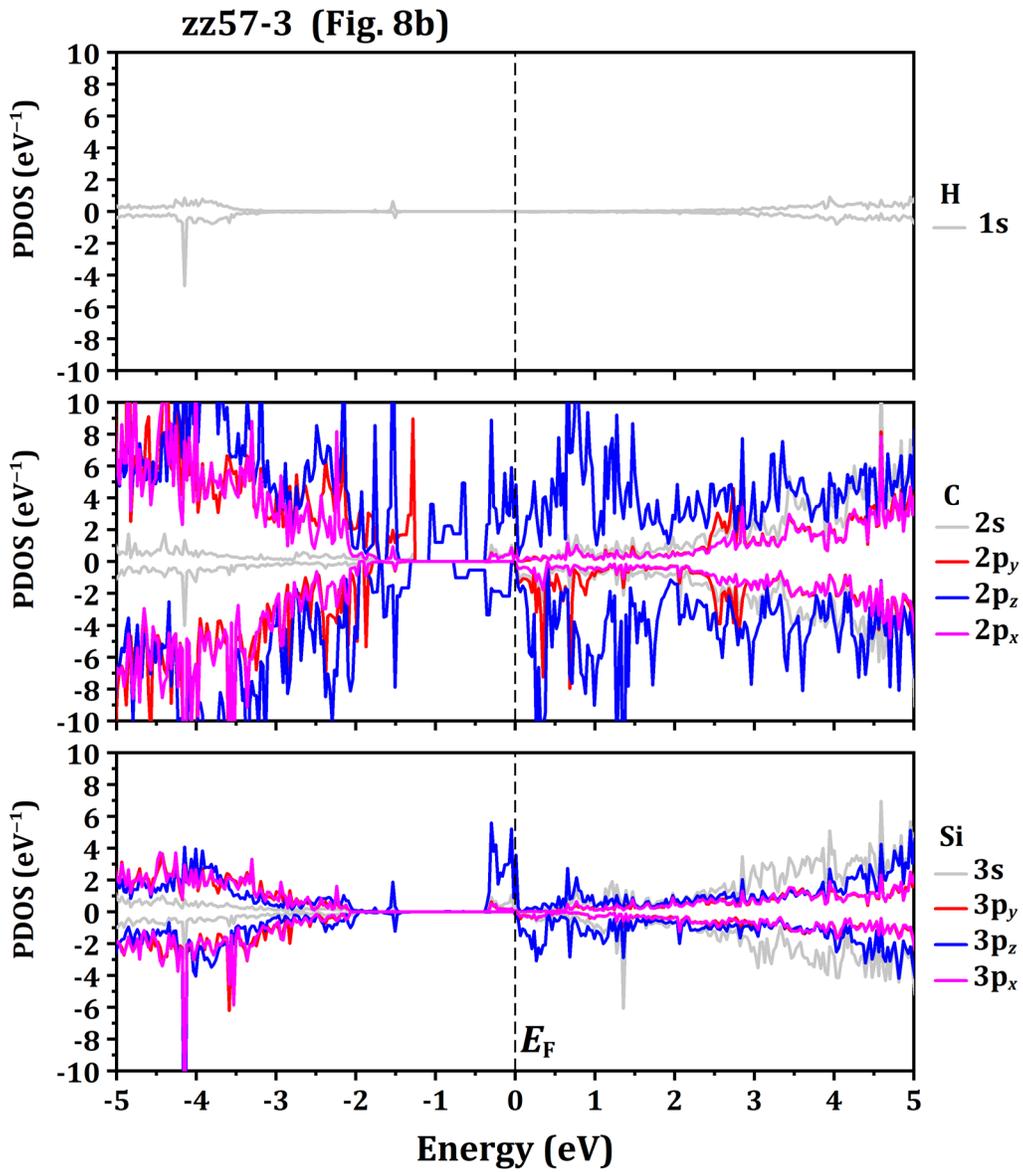

**FIG. S6.** The total PDOS of each element (H, C, or Si) for the configuration in Fig. 8(b) and also listed in Table S3.

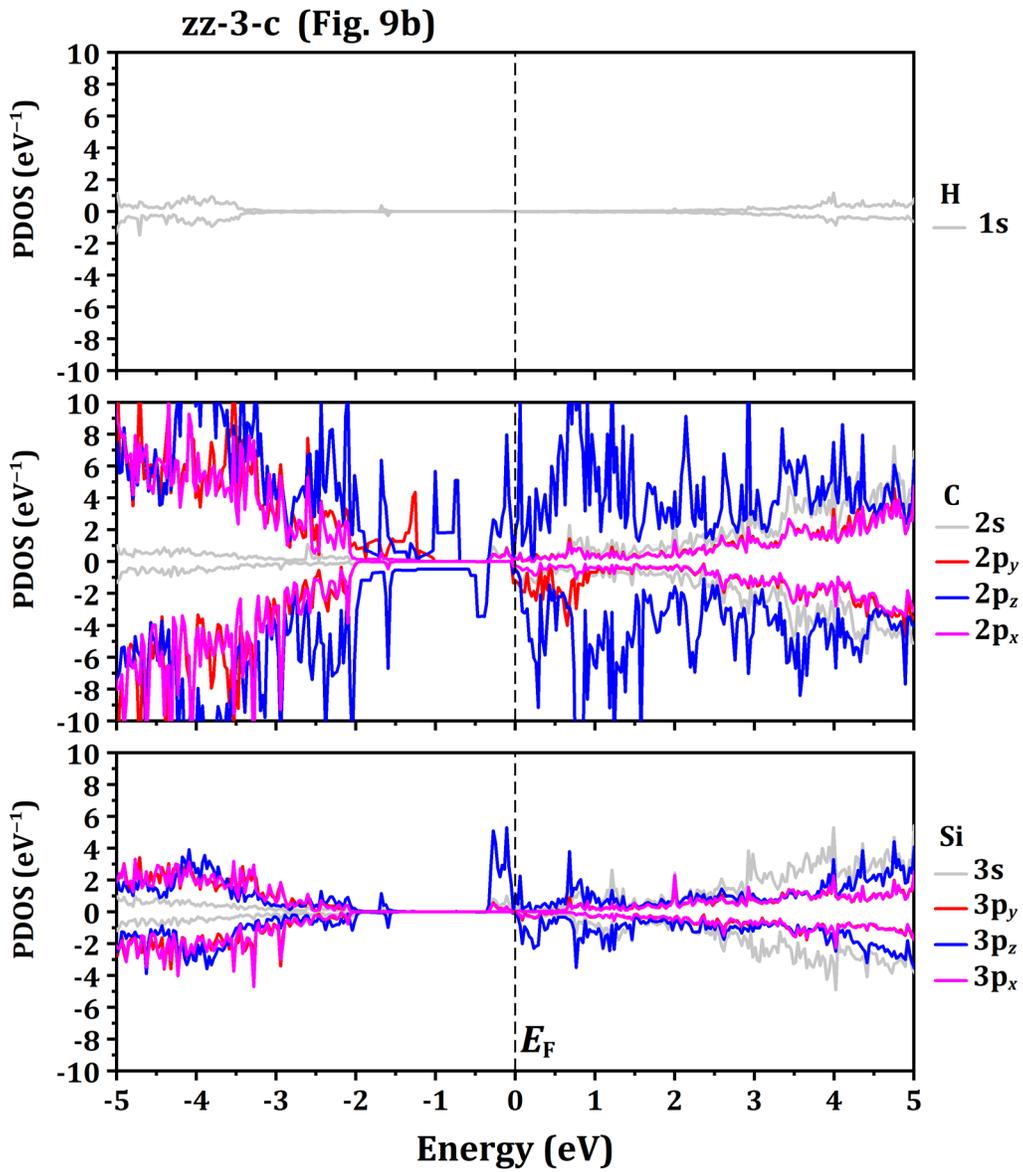

**FIG. S7.** The total PDOS of each element (H, C, or Si) for the configuration in Fig. 9(b) and also listed in Table S3.

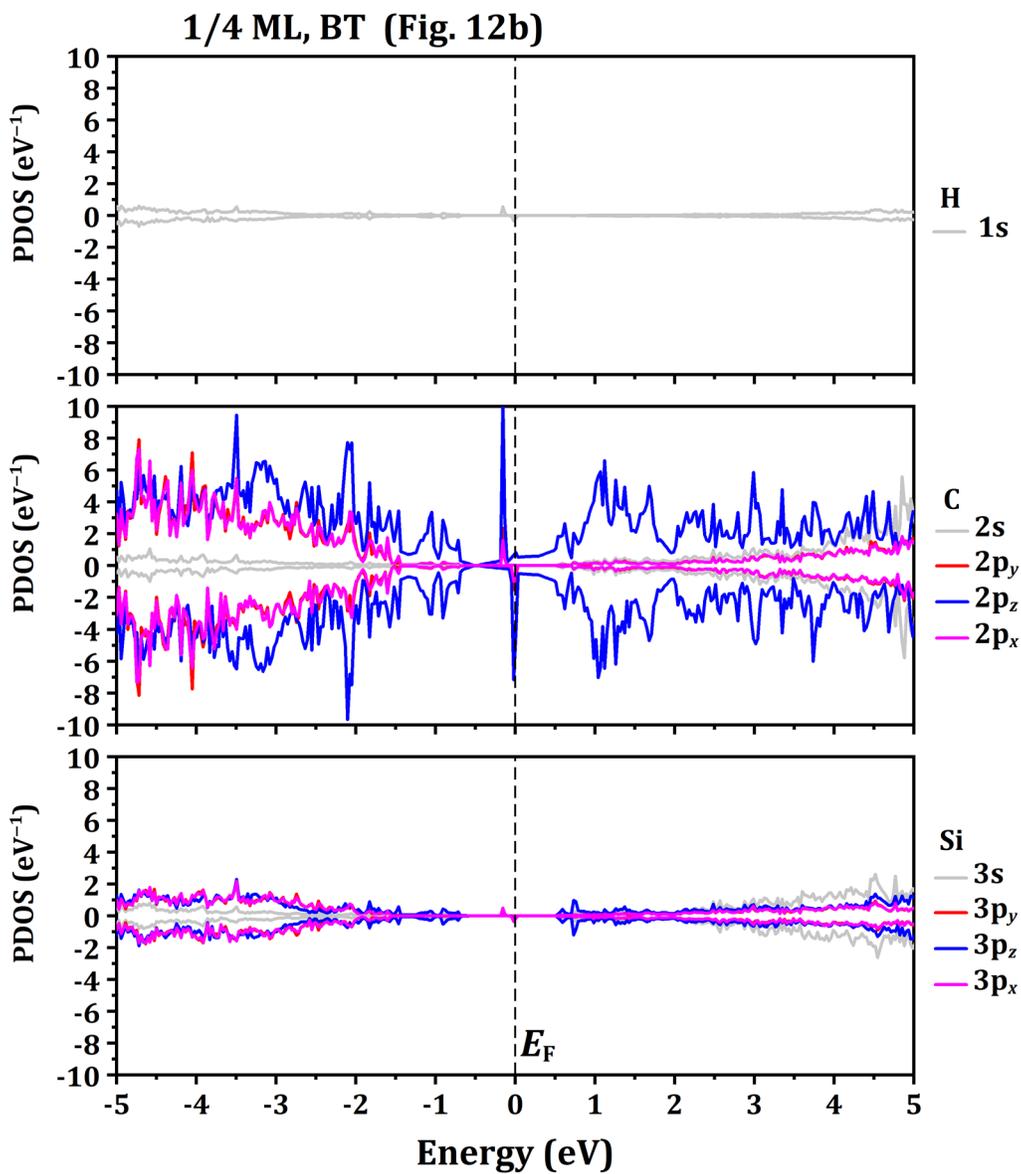

**FIG. S8.** The total PDOS of each element (H, C, or Si) for the configuration in Fig. 12(b) and also listed in Table S2.

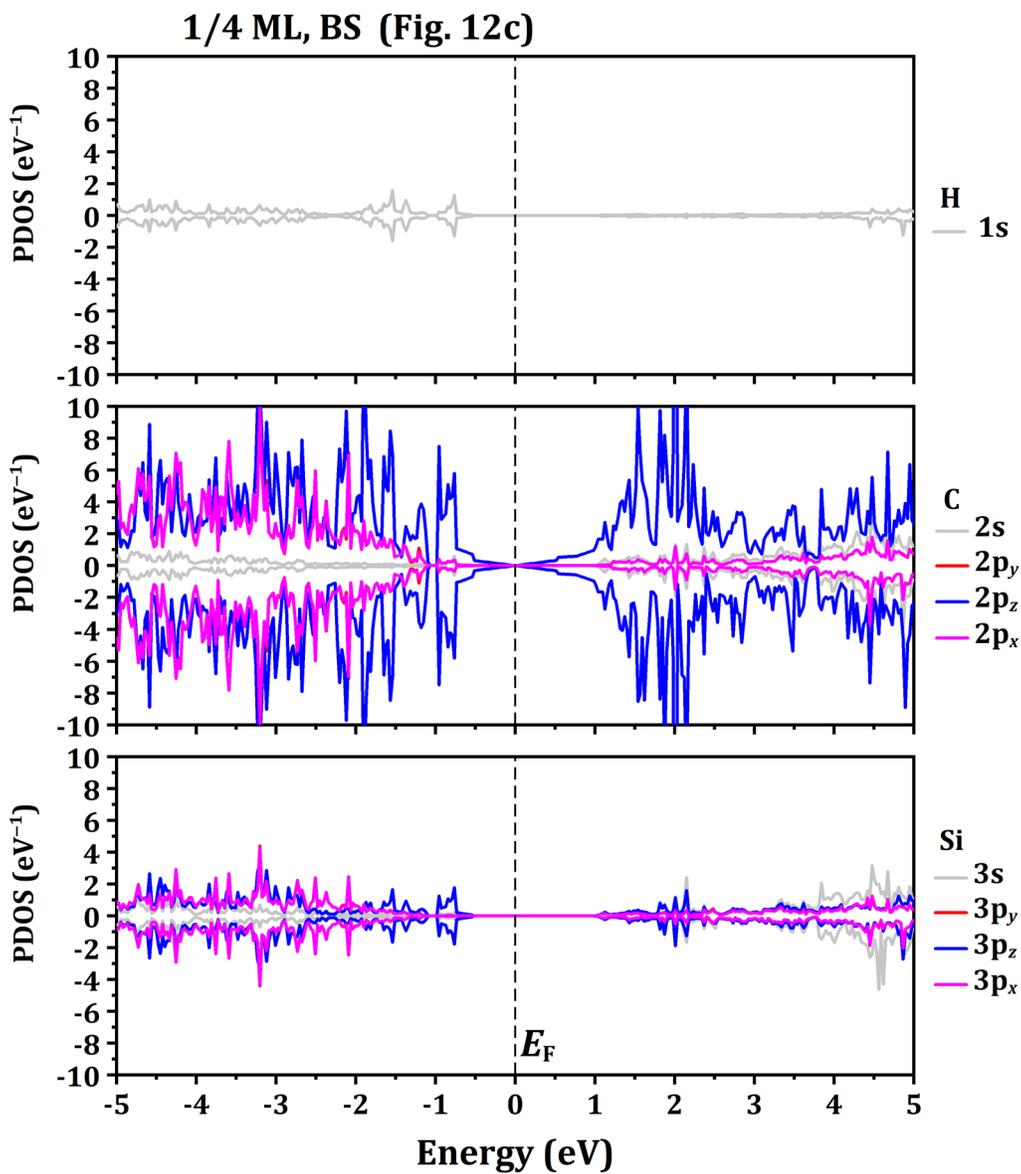

**FIG. S9.** The total PDOS of each element (H, C, or Si) for the configuration in Fig. 12(c) and also listed in Table S2.

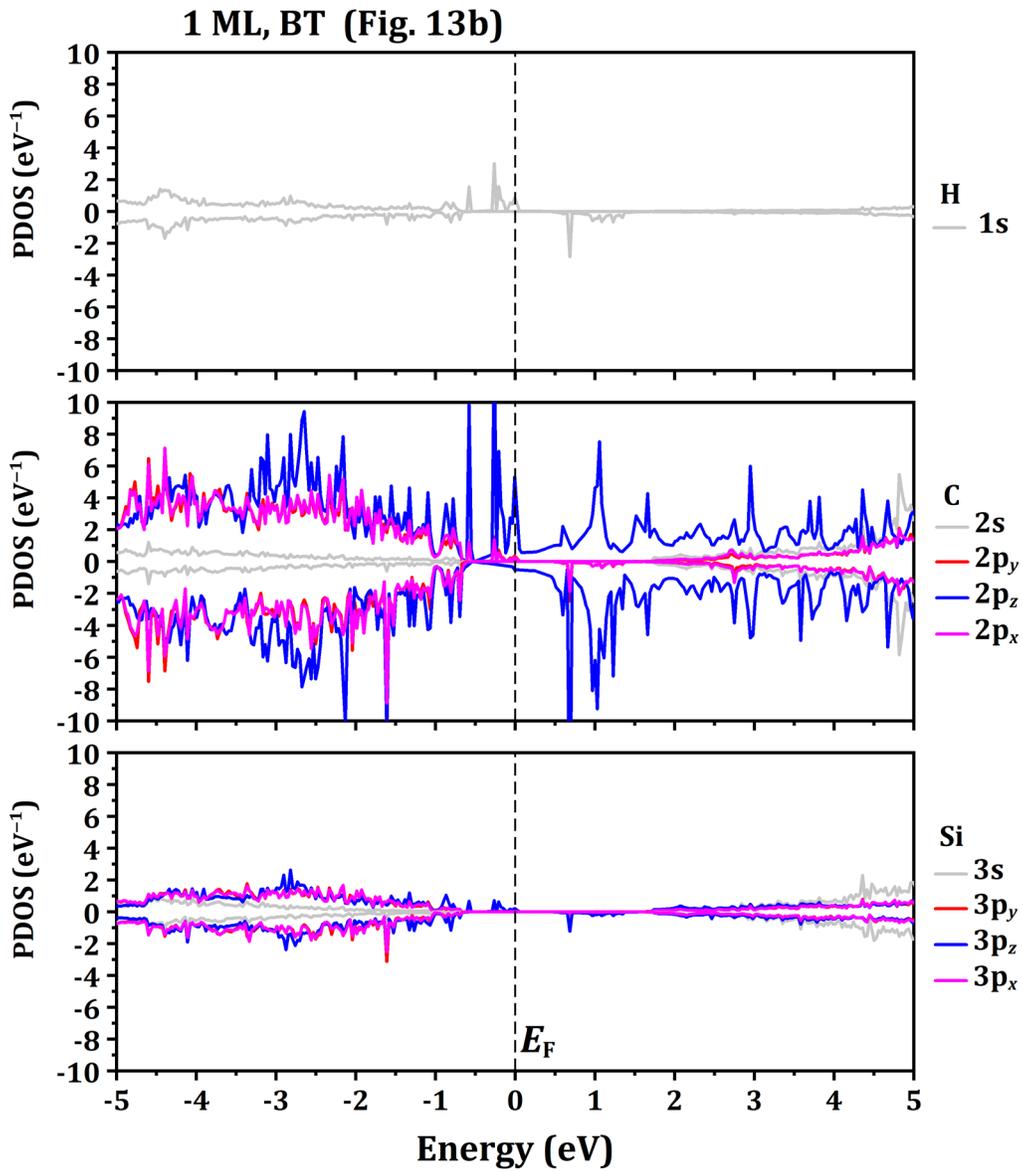

**FIG. S10.** The total PDOS of each element (H, C, or Si) for the configuration in Fig. 13(b) and also listed in Table S2.

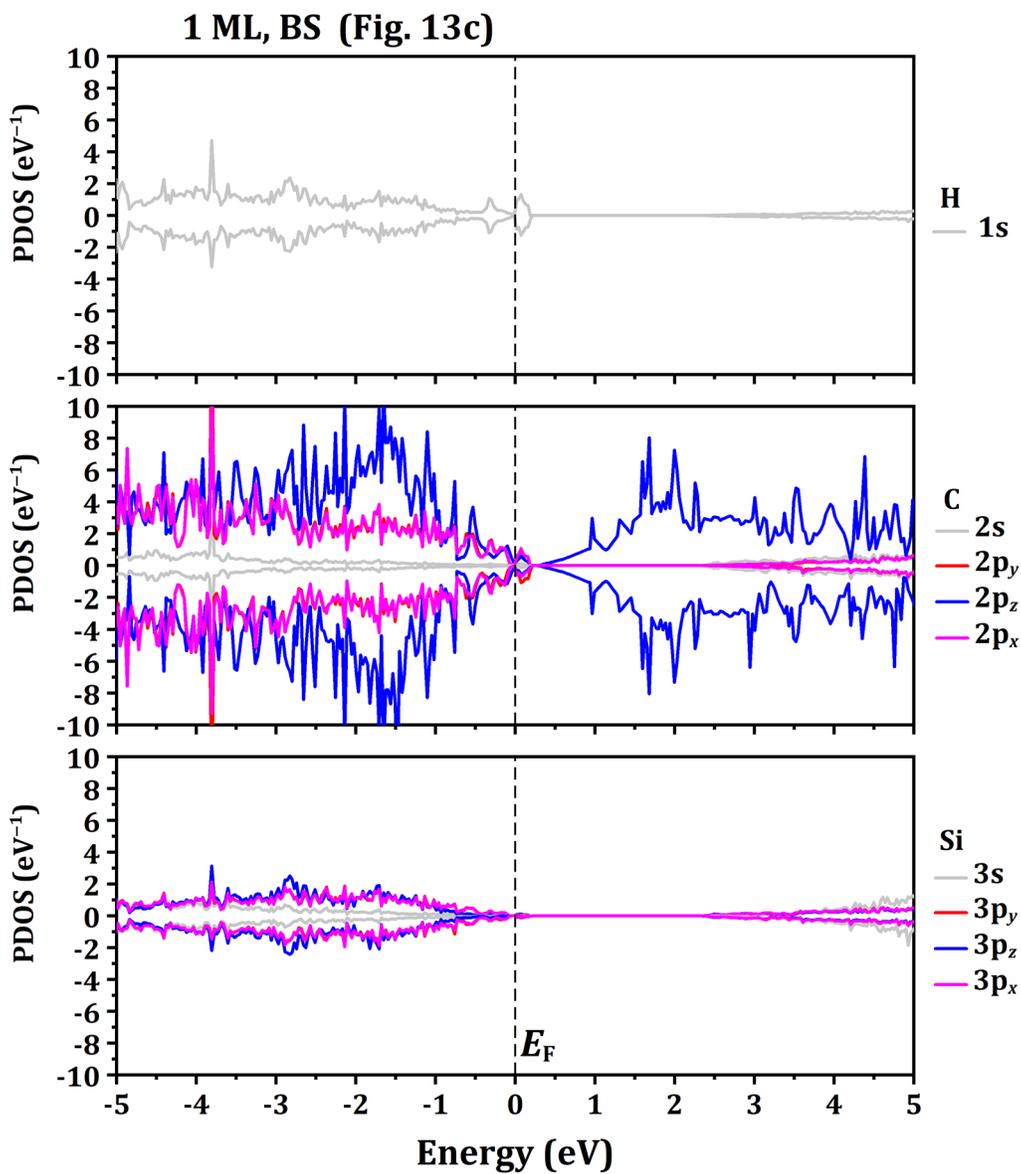

**FIG. S11.** The total PDOS of each element (H, C, or Si) for the configuration in Fig. 13(c) and also listed in Table S2.